\documentclass[fleqn,usenatbib]{mnras}
\usepackage{longtable}

\usepackage{newtxtext,newtxmath}

\usepackage[T1]{fontenc}

\DeclareRobustCommand{\VAN}[3]{#2}
\let\VANthebibliography\thebibliography
\def\thebibliography{\DeclareRobustCommand{\VAN}[3]{##3}\VANthebibliography}

\usepackage{booktabs} 
\usepackage[flushleft]{threeparttable}

\usepackage{graphicx}	
\usepackage{amsmath}	
\usepackage{xcolor}

\usepackage{tabularx}
\usepackage{hologo} 
\usepackage{booktabs} 

\newcommand{\rJ}{$r_{\rm J}$}

\newbox\grsign \setbox\grsign=\hbox{$>$} \newdimen\grdimen \grdimen=\ht\grsign
\newbox\simlessbox \newbox\simgreatbox
\setbox\simgreatbox=\hbox{\raise.5ex\hbox{$>$}\llap
     {\lower.5ex\hbox{$\sim$}}}\ht1=\grdimen\dp1=0pt
\setbox\simlessbox=\hbox{\raise.5ex\hbox{$<$}\llap
     {\lower.5ex\hbox{$\sim$}}}\ht2=\grdimen\dp2=0pt

\def\simless{\mathrel{\copy\simlessbox}}
\newbox\simppropto
\setbox\simppropto=\hbox{\raise.5ex\hbox{$\sim$}\llap
     {\lower.5ex\hbox{$\propto$}}}\ht2=\grdimen\dp2=0pt

\title[APOGEE globular clusters]{The APOGEE Value Added Catalogue of Galactic globular cluster stars}

\author[Schiavon et al.]{
Ricardo P. Schiavon,$^{1}$\thanks{E-mail: R.P.Schiavon@ljmu.ac.uk (RPS)}
Si\^an G. Phillips,$^{1}$
Natalie Myers$^{2}$,
Danny Horta$^{1,3}$,
Dante Minniti$^{4,5,6}$,
\newauthor
Carlos Allende Prieto$^{7,8}$,
Borja Anguiano$^{9,10}$,
Rachael L. Beaton$^{11,12}$,
Timothy C. Beers$^{13}$,
\newauthor
Joel R. Brownstein$^{14}$,
Roger E. Cohen$^{15}$,
Jos\'e G. Fern\'andez-Trincado$^{16}$,
Peter M. Frinchaboy$^{2}$,
\newauthor
Henrik J\"onsson$^{17}$,
Shobhit Kisku$^{1}$,
Richard R. Lane$^{18}$,
Steven R. Majewski$^{10}$,
Andrew C. Mason$^{1}$,
\newauthor
Szabolcs M\'esz\'aros$^{19,20}$,
Guy S. Stringfellow$^{21}$
\medskip
\\
$^{1}$ Astrophysics Research Institute, Liverpool John Moores University, 146 Brownlow Hill, Liverpool L3 5RF, UK\\
$^{2}$ Department of Physics \& Astronomy, Texas Christian University, Fort Worth, TX 76129, USA \\
$^{3}$ Center for Computational Astrophysics, Flatiron Institute, 162 5th Ave., New York, NY 10010, USA \\
$^{4}$ Instituto de Astrof\'\i sica, Facultad de Ciencias Exactas, Universidad Andr\'es Bello, Fern\'andez, Concha 700, Las Condes, Santiago, Chile \\
$^{5}$ Vatican Observatory, Vatican City State, V-00120, Italy \\
$^{6}$ Departamento de F\'\i sica, Universidade Federal de Santa Catarina, Trindade 88040-900, Florian\'opolis, Brazil \\
$^{7}$ Instituto de Astrof\'\i sica de Canarias, V\'\i a L\'actea S/N, 38205 La Laguna, Tenerife, Spain. \\
$^{8}$ Universidad de La Laguna, Departamento de Astrof\'\i sica, 38206 La Laguna, Tenerife, Spain \\
$^{9}$ Department of Physics and Astronomy, University of Notre Dame, 225 Nieuwland Science Hall, Notre Dame, IN 46556, USA \\
$^{10}$ Department of Astronomy, University of Virginia, Charlottesville, VA, 22904, USA \\
$^{11}$ Department of Astrophysical Sciences, 4 Ivy Lane, Princeton University, Princeton, NJ 08544 \\
$^{12}$ The Observatories of the Carnegie Institution for Science, 813 Santa Barbara St., Pasadena, CA 91101 \\
$^{13}$ Department of Physics and Astronomy and JINA Center for the Evolution of the Elements, University of Notre Dame, Notre Dame, IN 46556, USA \\
$^{14}$ Department of Physics and Astronomy, University of Utah, 115 S. 1400 E., Salt Lake City, UT 84112, USA \\
$^{15}$ Rutgers the State University of New Jersey, 136 Frelinghuysen Ave., Piscataway, NJ 08854, USA \\
$^{16}$ Instituto de Astronom\'ia, Universidad Cat\'olica del Norte, Av. Angamos 0610, Antofagasta, Chile\\
$^{17}$ Materials Science and Applied Mathematics, Malmö University,
SE-205 06 Malm\"o, Sweden \\
$^{18}$ Centro de Investigación en Astronomía, Universidad Bernardo O'Higgins, Avenida Viel 1497, Santiago, Chile \\
$^{19}$ ELTE E\"otv\"os Lor\'and University, Gothard Astrophysical Observatory, 9700 Szombathely, Szent Imre H. st. 112, Hungary \\
$^{20}$  MTA-ELTE Lend{\"u}let "Momentum" Milky Way Research Group, Hungary\\
$^{21}$ Center for Astrophysics and Space Astronomy, University of Colorado at Boulder, 389 UCB, Boulder, CO 80309-0389, USA
}

\date{Accepted XXX. Received YYY; in original form ZZZ}

\pubyear{2015}

\begin{document}
\label{firstpage}
\pagerange{\pageref{firstpage}--\pageref{lastpage}}
\maketitle

\begin{abstract}
We introduce the SDSS/APOGEE Value Added Catalogue of Galactic Globular Cluster (GC) Stars.  The catalogue is the result of a critical search of the APOGEE data release 17 (DR17) catalogue for candidate members of all known Galactic GCs.  Candidate members are assigned to various GCs on the basis of position on the sky, proper motion, and radial velocity.  
The catalogue contains a total of 7,737 entries for 6,422 unique stars associated with 72 Galactic GCs. Full APOGEE DR17 information is provided, including radial velocities and abundances for up to 20 elements.  Membership probabilities estimated on the basis of precision radial velocities are made available.  Comparisons with chemical compositions derived by the GALAH survey, as well as optical values from the literature, show good agreement.
This catalogue represents a significant increase in the public database of GC star chemical compositions and kinematics, providing a massive homogeneous data set that will enable a variety of studies.  The catalogue in fits format is available for public download from the SDSS-IV DR17 value added catalogue website.
\end{abstract}




\section{Introduction}

Globular clusters are intriguing objects.  The physics of their genesis is not entirely understood, yet their study has advanced knowledge in several fields of astrophysics.  The mapping of the spatial distribution of Galactic globular clusters (GCs) promoted a radical revision of the position of the solar system in the universe \citep{Shapley1918}; application of the physics of stellar structure and evolution to GC observations constrained the age of the universe in the early days of Big Bang cosmology \citep[e.g.,][and references therein]{Sandage1970,Bolte1995}; GC ages, chemical compositions, and orbital properties provide important clues to the star formation and accretion history of the early Milky Way \citep[e.g.,][]{SearleZinn1978,SalarisWeiss2002,Massari2019,Kruijssen2019,Horta2020,Forbes2020,Callingham2022} as well as other galaxies \citep[][and references therein]{BrodieStrader2006}; finally and no less crucially, to this day GCs are fundamental test beds of stellar evolution theory in the low mass regime \citep[e.g.,][]{Schwarzschild1970,Renzini1988,Chiosi1992,Salaris2002}.  Yet after over a century of study, their origin is still subject to debate, with no shortage of formation scenarios \citep[e.g.,][]{FallRees1985,Schweizer1987,AshmanZepf1992} despite recent encouraging progress \citep[e.g.,][]{Kruijssen2015,Choksi2018,Pfeffer2018}.

Since GCs stand at the crossroads of many areas of astrophysics, it is small wonder that they have been subject to various herculean observational efforts over the past several decades---in fact so many that an exhaustive account is rendered impossible in this brief introduction. We thus limit ourselves to mention a few highlights and some of the most recent work, in a manner dictated by the authors' own personal biases and an unavoidably limited grasp of an overwhelming---and ever growing---literature.  

Systematic photometric observations built ground and space based colour-magnitude diagrams of large GC samples in the optical \citep[e.g.,][]{Barbuy1998,Rosenberg2000,Piotto2002,Sarajedini2007,Stetson2019}, near infrared \citep[e.g.,][]{Cohen2017,Minniti2017}, and ultraviolet \cite[e.g.,][]{Schiavon2012,Sahu2022}.  Libraries of integrated spectra were created for comparison against observations of extragalactic GCs and reality checking of stellar population synthesis models \citep[e.g.,][]{ZinnWest1984,BicaAlloin1986,ArmandroffZinn1988,Puzia2002,Schiavon2005}, and more recently integral field spectroscopy of large GC samples have also become available \citep[e.g.,][]{Usher2017,Kamann2018}. In this context, the GC system of the Andromeda galaxy has also become subject to extensive integrated light surveys both in optical and NIR \citep[e.g.,][]{Galleti2007,Caldwell2009,Schiavon2013,Sakari2016}. Finally, with the advent of the {\it Gaia} satellite, a massive undertaking by E.~Vasiliev and H.~Baumgardt has produced precision kinematics and structural parameters for the vast majority of known Galactic GCs \citep{Baumgardt2021,Vasiliev2021}.

Chemical compositions of individual GC stars constitute precious, and quite expensive, information for a variety of scientific pursuits. Within the confines of stellar evolution theory, such pursuits include the calibration of evolutionary tracks, the study of stellar evolution processes such as dredge up and deep mixing along the giant branch \citep[e.g.,][]{Kraft1979,Shetrone1996}, and diffusion of heavy elements in main sequence stars \citep[e.g.,][]{DenissenkovWeiss1996,Castellani1999,Lind2008}. The first systematic collection of chemical compositions of individual stars in Galactic GCs was conducted by the Lick/Texas group \citep[e.g.,][]{Kraft1994,Shetrone1996,Sneden1997}.  Contrary to the generally agreed notion of GCs as coeval stellar systems with homogeneous chemical compositions, these early efforts revealed that star-to-star abundance variations are ubiquitous.  These were difficult to understand, but since the data were restricted to bright giant stars, the broad consensus was that such variations should be ascribed to stellar evolution effects.

The next generation of systematic measurements brought about a considerable amplification of the existing database of homogeneously derived chemical compositions \cite[e.g.,][and references therein]{CG97,Carretta2010}.  These efforts consolidated the knowledge that star-to-star chemical composition variations are the norm in GCs.  They typically manifest themselves in the form of anti-correlations between the abundances of light elements such as C-N, Na-O, and Mg-Al \citep[e.g.,][]{Gratton2012}. Such abundance variations are present in main sequence stars \citep[e.g.,][]{Cannon1998}, ruling out evolutionary effects as their physical origin.  In addition to these features, massive systems such as $\omega$~Cen\footnote{It bears mentioning at this point that $\omega$~Cen is now believed to be the remnant nuclear cluster of a dwarf galaxy long accreted to the Milky Way \cite[e.g.,][]{Bekki2003,Majewski2012}, which more recently has been potentially identified \citep[e.g.,][]{Massari2019,Pfeffer2021} as {\it Gaia} Enceladus/Sausage \citep[][]{Belokurov2018,Helmi2018} or Sequoia \citep[][]{Myeong2019,Forbes2020}.}, among others, display variations in the abundances of the heavy elements that are by-products of SN~Ia enrichment \citep[e.g.,][]{Pancino2002,Johnson2010}. 

The discovery of multiple sequences in colour-magnitude diagrams, made possible by the
significant increase in photometric precision afforded by the HST/ACS, 
prompted the conclusion that GCs host a complex mix of stellar populations. In view of this overwhelming evidence, the historical assumption that GCs are single stellar populations had to be dropped.  This so called {\it multiple populations phenomenon}, is without a doubt inextricably linked to the physics of GC formation. Yet no formation scenarios are capable of accounting for this phenomemon in a quantitative fashion \citep[see reviews by][]{Renzini2015,BastianLardo2018,Milone2022}.

A solution to the problem of multiple populations in GCs, and in a broader perspective our understanding of the nature of these beautiful and fascinating systems, can be advanced by the production of a massive, homogeneous, and publicly available database of chemical compositions and kinematics for a large sample of Galactic GCs.  This paper summarises the effort by the APOGEE\footnote{Apache Point Observatory Galactic Evolution Experiment} team to make one such database available to public access.  We present the APOGEE value added catalogue (VAC) of Galactic globular cluster members. The paper is organised as follows. In Section~\ref{sec:data} we briefly describe the APOGEE data.  In Section~\ref{sec:selection} the criteria adopted for selecting candidate GC members are described, while membership probabilities are discussed in Section~\ref{sec:probs}.  Broad features of the data are presented in Section~\ref{sec:results}, and Section~\ref{sec:summary} describes the catalogue and provides access information.

\section{The data}
\label{sec:data}

This paper presents a catalogue of chemical compositions and radial velocities from the latest data release of the SDSS-IV/APOGEE 2 survey \citep[DR17,][]{Majewski2017,sdss_dr17}.  Proper motions from the early third data release (eDR3) from the {\it Gaia} satellite \citep{Gaia_eDR3} and various additional metadata imported directly from the DR17 catalogue are also included for convenience sake.  Data from APOGEE have been described in detail in various technical publications, so we provide a brief account of their main properties in this Section, referring the reader to the relevant papers for further details.  Chemical-composition data based on earlier APOGEE data releases were presented for various collections of Galactic GCs in a number of publications \citep[e.g.,][]{Meszaros2015,Schiavon2017b,Nataf2019,Masseron2019,Meszaros2020,Meszaros2021,Geisler2021}.

Elemental abundances and radial velocities are obtained from the automatic analysis of moderately high-resolution near-infrared spectra of hundreds of thousands of stars observed with the Apache Point Observatory 2.5~m Sloan telescope \citep{Gunn2006} and the Las Campanas Observatory 2.5~m Du Pont telescope \cite{BowenVaughan1973}.  The telescopes are equipped with twin high efficiency multi-fiber NIR spectrographs designed and assembled at the University of Virginia, USA \citep{Wilson2019}. A technical summary of the overall SDSS-IV experiment can be found in \cite{Blanton_2017}. 

\begin{figure}
	\includegraphics[width=\columnwidth]{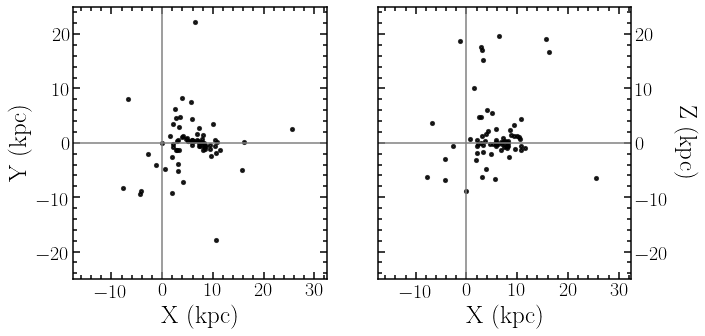}
    \caption{Spatial distribution of the GCs included in this VAC, in Cartesian coordinates with the Sun at the origin (cross hairs). The sample is preferentially concentrated towards the inner Galaxy.}
    \label{fig:spatial}
\end{figure}

APOGEE spectra for any given star (save for a relatively small number of exceptions) were obtained in a number of visits separated in time to enable the detection of radial velocity variations caused by binarity.  Observations spanned a period of typically 3 months, never exceeding 6 months between first and last visit.  Every visit spectrum was integrated typically for $\sim$1 hour in sets of four (ABBA) exposures taken with the detector dithered along the spectral direction.  Spectral dithers were aimed at bringing the sampling of the line spread function to slightly better than critical in the blue end of the spectrum where the sampling by the detector's original pixel size is sub-critical.  

A pipeline built specifically for the reduction of APOGEE data \citep{Nidever2015} was employed to apply standard operations such as reference pixel voltage correction, linearisation, cosmic-ray and saturation corrections, dark current subtraction, persistence correction, and extraction of the 2D data array from the 3D data cubes.  The 2D images were then flat-field corrected and a bad-pixel mask was generated before 1D spectra were extracted, and subsequently wavelength and flux calibrated.  The next step was to subtract the sky background, which is dominated by emission lines, and perform telluric correction before combining the dithered sequences into a single well sampled resulting visit spectrum.  

For each star, relative radial velocities of visit spectra were measured through iterative cross correlation with the combined spectrum, and final absolute RVs are then obtained by cross-correlation of the combined spectrum with a grid of synthetic spectra covering a wide range of stellar parameters.  The resulting combined restframe spectrum is that which is finally fed into the stellar abundances pipeline.  

The APOGEE Stellar Parameters and Chemical Abundances Pipeline (ASPCAP) is described in detail by \cite{GarciaPerez2016} and later updates  \citep{Holtzman2018,Jonsson2020}.  In short, it determines stellar parameters and detailed elemental abundances through interpolation, using the FERRE\footnote{Available at https://github.com/callendeprieto/ferre} software \citep{AllendePrieto2006}, into a huge grid of synthetic spectra covering the entire range of stellar parameters and chemical compositions of interest.  The spectral library is calculated adopting MARCS model atmospheres \citep{Gustafsson2008} generated specifically for the purposes of the APOGEE survey \citep[see][]{Meszaros2012,Zamora2015,Holtzman2018,Jonsson2018}.  ASPCAP abundance analysis is based on various flavours of line lists, depending on the spectrum synthesis codes, LTE/non-LTE assumption, and model atmospheres adopted in the construction of the spectral library.  Those line lists are empirically tuned to match the observed high-resolution spectra of the Sun and Arcturus.  For details, see  \cite{Smith2021} \citep[but see also][]{Cunha2017,Hasselquist2016,Shetrone2015}.  

The data described above were made publicly available in data release 17 in the form of various catalogues for different flavours of spectral analysis, according to the spectrum synthesis code adopted in the construction of the spectral library, adoption or not of an NLTE approach for some elements\footnote{For details see https://www.sdss.org/dr17/irspec/spectro\_data\_supplement}, and the assumption of plane-parallel or spherical geometry in the radiative transfer calculation.  Each catalogue contains 733,901 entries.  The data contained in this VAC are extracted from the default DR17 data analysis, which is based on the Synspec-based \citep{HubenyLanz2017} spectral library, with incorporation of an NLTE abundance analysis for elements Na, Mg, K, and Ca \citep[prefix {\tt synspec\_rev1}, see][and references therein]{Osorio2020}.

\subsection{Globular Cluster Sample} \label{sec:sample}

Globular clusters were targeted by APOGEE so as to satisfy at the same time scientific interests and calibration needs.  As the first attempt at automatic detailed chemical composition analysis of a massive near infrared spectroscopic database, optical calibrators are a crucial requirement for APOGEE.  As targets of interest for various scientific pursuits, GC stars have for decades been the focus of chemical composition studies.  Thus the availability of multiple elemental abundance determinations in the literature, overwhelmingly based on optical spectroscopy, placed GC stars at the centre of the APOGEE calibration procedure \citep[e.g.,][]{Holtzman2015,Holtzman2018,Jonsson2020}.  With those goals in mind, a large number of GCs were targeted during the execution of both APOGEE~1 and 2.  Targets within each GC were selected to meet a set of criteria whereby stars that were subject to previous abundance analysis and/or atmospheric parameter determinations were given top priority, followed by stars with membership confirmed on the basis of radial velocity, proper motion, and position on the colour-magnitude diagram, in decreasing order of priority \citep[for details, see][]{Zasowski2013,Zasowski2017,Santana2021,Beaton2021}.  In addition to the GCs targeted as part of the main APOGEE survey, a number of additional systems were observed as part of the bulge Cluster APOgee Survey \citep[CAPOS,][]{Geisler2021}.  The CAPOS team took advantage of Chilean access to the APOGEE South spectrograph \citep{Wilson2019} to collect data for a number of GCs located towards the inner Galaxy. CAPOS spectra were collected, reduced and analysed following the same procedures and pipelines as the main survey targets, with results being ingested into the SDSS/APOGEE database.  Data obtained by CAPOS are thus treated in this paper in the same way as the targets from the main APOGEE survey.  Table 2 lists the GCs included in this value added catalogue, along with basic parameters, extracted from \cite{Baumgardt2021,Vasiliev2021}\footnote{Available at https://people.smp.uq.edu.au/HolgerBaumgardt/globular/}, which we hereafter refer simply as the VB catalogue.  The number of candidate members of each GC is also listed.  The distribution of our GC sample in Cartesian coordinates is shown in Figure~\ref{fig:spatial}. Unfortunately, GCs whose discovery was reported after the latest update of the VB catalogue, such as VVV CL001 \citep{FT2021} and Patchick 125 \citep{FT2022} are not included in this catalogue, but will be incorporated in future versions.

\begin{table*} 
\label{tab:list}
	\centering
	\caption{Globular clusters included in the sample. Column information: (1) GC ID; (2,3) coordinates of GC centre; (4) mean iron abundance; (5) mean radial velocity; (6) heliocentric distance; (7) Galactocentric distance; (8) mass; (9) Jacobi radius; (10) number of entries. Numbers for columns 2, 3, 5, 6, 7, 8, and 9 are from the VB catalogue, whereas those for columns 4 and 10 are from this work.}
	\begin{tabular}{lrrccrrccr} 
		\hline
		\hline
GC  &  $\alpha_{\rm cen}$ & $\delta_{\rm cen}$  &  $\langle{\rm \lceil Fe/H \rceil}\rangle$ &  $\langle{\rm R.V.}\rangle$ & $d_\odot$  &  $R_{\rm GC}$  &  Mass  &  $r_{\rm J} $ &  N\\
		\hline
 &  deg  & deg  &  & km~s$^{-1}$ & kpc & kpc & 10$^4$~M$_\odot$ & deg & \\
		\hline
  NGC~104  & 6.02379    & -72.08131  & -0.74   &  -17.45 $\pm$  0.16    &   4.52  $\pm$  0.03  &  7.52  $\pm$ 0.01   &   89.5   $\pm$ 0.6   & 1.557  & 297      \\
  NGC~288  & 13.18850    & -26.58261  & -1.27   &  -44.45 $\pm$  0.13    &   8.99  $\pm$  0.09  &  12.21 $\pm$ 0.06   &   9.3  $\pm$ 0.3   & 0.605  & 43       \\
  NGC~362  & 15.80942   & -70.84878  & -1.11   &  223.12 $\pm$  0.28    &   8.83  $\pm$  0.10  &  9.62  $\pm$ 0.06   &   28.4   $\pm$ 0.4   & 0.598  & 70       \\
  Palomar~1     & 53.33350    & 79.58105   & -0.45   &  -75.72 $\pm$  0.29    &   11.28 $\pm$  0.32  &  17.41 $\pm$ 0.29   &   0.10 $\pm$ 0.02  & 0.122  & 3        \\
  NGC~1851  & 78.52816   & -40.04655  & -1.13    &  321.4  $\pm$  1.55   &   11.95 $\pm$  0.13  &  16.69 $\pm$ 0.11   &   31.8   $\pm$ 0.4   & 0.611  & 71       \\
  NGC~1904  & 81.04584   & -24.52442  & -1.52   &  205.76 $\pm$  0.2     &  13.08 $\pm$  0.18  &  19.09 $\pm$ 0.16   &   13.9   $\pm$ 1.0    & 0.259  & 40       \\
  NGC~2298  & 102.24754  & -36.00531  & -1.84   &  147.15 $\pm$  0.57    &   9.83 $\pm$  0.17  &  15.07 $\pm$ 0.14   &   5.6   $\pm$ 0.8 & 0.435  & 12       \\
  NGC~2808  & 138.01291  & -64.86349  & -1.07   &  103.57 $\pm$  0.27    &  10.06 $\pm$  0.11  &  11.58 $\pm$ 0.07   &   86.4   $\pm$ 0.6    & 0.944  & 132      \\
  NGC~3201  & 154.40343  & -46.41248  & -1.39   &  493.65 $\pm$  0.21    &   4.74 $\pm$  0.04  &  8.93  $\pm$ 0.02   &   16.0   $\pm$ 0.3   & 0.925  & 217      \\
  NGC~4147  & 182.52626  & 18.54264   & -1.63   &  179.35 $\pm$  0.31    &  18.53 $\pm$  0.21  &  20.2 $\pm$ 0.2   &   3.9   $\pm$ 0.9   & 0.274  & 3        \\
  Rup~106   & 189.66750   & -51.15028  & -1.30   &  -38.36 $\pm$  0.26    &  20.71 $\pm$  0.36  &  18.0 $\pm$ 0.3   & 3.4  $\pm$ 0.6    & 0.213  & 2        \\
  NGC~4590  & 189.86658  & -26.74406  & -2.22   &  -93.11 $\pm$  0.18    &  10.40 $\pm$  0.10  &  10.35 $\pm$ 0.07   &   12.2   $\pm$ 0.9   & 0.426  & 41       \\
  NGC~5024  & 198.23021  & 18.16817   & -1.90   &  -63.37 $\pm$  0.25    &  18.50 $\pm$  0.18  &  19.0  $\pm$ 0.16   &   45.5   $\pm$ 3.0    & 0.549  & 41       \\
  NGC~5053  & 199.11288  & 17.70025   & -2.21   &  42.82  $\pm$  0.25    &  17.54 $\pm$  0.23  &  18.01 $\pm$ 0.2    &   7.4  $\pm$ 2.0    & 0.317  & 17       \\
  NGC~5139  & 201.69699  & -47.47947  & -1.60   &  232.78 $\pm$  0.21    &   5.43 $\pm$  0.05  &  6.5   $\pm$ 0.01   &   364   $\pm$ 4    & 2.142  & 1864     \\
  NGC~5272  & 205.54842  & 28.37728   & -1.43   &  -147.2 $\pm$  0.27    &  10.17 $\pm$  0.08  &  12.09 $\pm$ 0.06   &   41.0   $\pm$ 1.7     & 0.714  & 299      \\
  NGC~5466  & 211.36371  & 28.53444   & -1.81   &  106.82 $\pm$  0.2     &  16.12 $\pm$  0.16  &  16.47 $\pm$ 0.13   &   6.0  $\pm$ 1.0   & 0.284  & 17       \\
  NGC~5634  & 217.40533  & -5.97643   & -1.72    &  -16.07 $\pm$  0.6     &  25.96 $\pm$  0.62  &  21.84 $\pm$ 0.57   &   22.8   $\pm$ 4.0    & 0.42   & 2        \\
  Palomar~5     & 229.01917  & -0.121     & -1.24   &  -58.61 $\pm$  0.15    &   21.94 $\pm$  0.51  &  17.27 $\pm$ 0.47   &   1.0 $\pm$ 0.2   & 0.12   & 12       \\
  NGC~5904  & 229.63841  & 2.08103    & -1.21   &  53.5   $\pm$  0.25    &   7.48 $\pm$  0.06  &  6.27  $\pm$ 0.02   &   39.4   $\pm$ 0.6   & 0.607  & 259      \\
  NGC~6093  & 244.26004  & -22.97608  & -1.61   &  10.93  $\pm$  0.39    &  10.34 $\pm$  0.12  &  3.95  $\pm$ 0.08   &   33.8   $\pm$ 0.9   & 0.344  & 3        \\
  NGC~6121  & 245.86974  & -26.52575  & -1.07   &  71.21  $\pm$  0.15    &   1.85 $\pm$  0.02  &  6.45  $\pm$ 0.01   &   8.7  $\pm$ 0.1   & 1.658  & 224      \\
  NGC~6144  & 246.80777  & -26.0235   & -1.80   &  194.79 $\pm$  0.58    &   8.15 $\pm$  0.13  &  2.5   $\pm$ 0.02   &   7.9  $\pm$ 1.4    & 0.198  & 1        \\
  NGC~6171  & 248.13275  & -13.05378  & -1.02   &  -34.71 $\pm$  0.18    &   5.63 $\pm$  0.08  &  3.74  $\pm$ 0.04   &   7.5  $\pm$ 0.4   & 0.368  & 65       \\
  NGC~6205  & 250.42181  & 36.45986   & -1.48   &  -244.9 $\pm$  0.3     &   7.42 $\pm$  0.08  &  8.64  $\pm$ 0.04   &   54.5   $\pm$ 2.0    & 1.036  & 152      \\
  NGC~6229  & 251.74525  & 47.5278    & -1.24   &  -137.89$\pm$  0.71    &  30.11 $\pm$  0.47  &  29.45 $\pm$ 0.44   &   28.6   $\pm$ 9.0    & 0.395  & 11       \\
  NGC~6218  & 251.80907  & -1.94853   & -1.27    &  -41.67 $\pm$  0.14    &   5.11 $\pm$  0.05  &  4.57  $\pm$ 0.02   &   10.7   $\pm$ 0.3   & 0.508  & 107      \\
  NGC~6254  & 254.28772  & -4.10031   & -1.51   &  74.21  $\pm$  0.23    &   5.07 $\pm$  0.06  &  4.35  $\pm$ 0.03   &   20.5   $\pm$ 0.4    & 0.611  & 87       \\
  NGC~6273  & 255.65749  & -26.26797  & -1.71    &  145.54 $\pm$  0.59    &   8.34 $\pm$  0.16  &  1.43  $\pm$ 0.03   &   69.7   $\pm$ 3.6    & 0.266  & 81       \\
  NGC~6293  & 257.54250   & -26.58208  & -2.09    &  -143.66$\pm$  0.39    &   9.19 $\pm$  0.28  &  1.6   $\pm$ 0.18   &   20.5   $\pm$ 1.6    & 0.153  & 20       \\
  NGC~6304  & 258.63440   & -29.46203  & -0.48   &  -108.62$\pm$  0.39    &   6.15 $\pm$  0.15  &  2.19  $\pm$ 0.13   &   12.6   $\pm$ 1.1    & 0.276  & 34       \\
  NGC~6316  & 259.15542  & -28.14011  & -0.77   &  99.65  $\pm$  0.84    &  11.15 $\pm$  0.39  &  3.16  $\pm$ 0.36   &   32.8   $\pm$ 4.0    & 0.271  & 24       \\
  NGC~6341  & 259.28076  & 43.13594   & -2.25   &  -120.55$\pm$  0.27    &   8.50 $\pm$  0.07  &  9.84  $\pm$ 0.04   &   35.2   $\pm$ 0.4   & 0.808  & 80       \\
  Terzan~2     & 261.88792  & -30.80233  & -0.86   &  134.56 $\pm$  0.96     &   7.75 $\pm$  0.33  &  0.74  $\pm$ 0.16   &   13.6  $\pm$ 2.5    & 0.084  & 5        \\
  Terzan~4     & 262.66251  & -31.59553  & -1.38   &  -48.96 $\pm$  1.57    &    7.59 $\pm$  0.31  &  0.82  $\pm$ 0.2    &   20.0   $\pm$ 5.0    & 0.125  & 3        \\
  HP~1      & 262.77167  & -29.98167  & -1.21   &  39.76  $\pm$  1.22    &    6.99 $\pm$  0.14  &  1.26  $\pm$ 0.13   &   12.4   $\pm$ 1.7    & 0.156  & 17       \\
  FSR~1758  & 262.8      & -39.808    & -1.42   &  227.31 $\pm$  0.59    &  11.08 $\pm$  0.74  &  3.46  $\pm$ 0.63   &   62.8   $\pm$ 5.6    & 0.618  & 15       \\
  Liller~1  & 263.35233  & -33.38956  & -0.14   &  60.36  $\pm$  2.44    &   8.06 $\pm$  0.35  &  0.74  $\pm$ 0.07   &   91.5   $\pm$ 14.7     & 0.261  & 30       \\
  NGC~6380  & 263.61861  & -39.06953  & -0.78   &  -1.48  $\pm$  0.73    &   9.61 $\pm$  0.30  &  2.15  $\pm$ 0.21   &   33.4   $\pm$ 0.5   & 0.233  & 28       \\
  Ton~2     & 264.03929  & -38.54092  & -0.74   &  -184.72$\pm$  1.12    &   6.99 $\pm$  0.33  &  1.76  $\pm$ 0.19   &   6.9  $\pm$ 1.6     & 0.166  & 11       \\
  NGC~6388  & 264.07178  & -44.7355   & -0.49   &  83.11  $\pm$  0.45    &  11.17 $\pm$  0.16  &  3.99  $\pm$ 0.13   &   125.0   $\pm$ 1.0    & 0.516  & 75       \\
  NGC~6401  & 264.65219  & -23.9096   & -1.09   &  -105.44$\pm$  2.5     &   8.06 $\pm$  0.24  &  0.75  $\pm$ 0.04   &   14.5   $\pm$ 0.2   & 0.094  & 7        \\
  NGC~6397  & 265.17538  & -53.67434  & -2.02   &  18.51  $\pm$  0.08    &   2.48 $\pm$  0.02  &  6.01  $\pm$ 0.02   &   9.7  $\pm$ 0.1   & 1.174  & 187      \\
  Palomar~6     & 265.92581  & -26.22499  & -0.92   &  177.0  $\pm$  1.35     &   7.05 $\pm$  0.45  &  1.33  $\pm$ 0.45   &   9.5  $\pm$ 1.7   & 0.124  & 6        \\
  Terzan~5     & 267.02020   & -24.77906  & -0.78   &  -82.57 $\pm$  0.73     &   6.62 $\pm$  0.15  &  1.65  $\pm$ 0.13   &   93.5   $\pm$ 6.9    & 0.422  & 24       \\
  NGC~6441  & 267.55441  & -37.05145  & -0.49   &  18.47  $\pm$  0.56    &  12.73 $\pm$  0.16  &  4.78  $\pm$ 0.15   &   132.0   $\pm$ 1.0     & 0.502  & 25       \\
  UKS1     & 268.61331  & -24.14528  & -1.00     &  59.38  $\pm$  2.63    &   15.58 $\pm$  0.56  &  7.7  $\pm$ 0.5   &   7.7   $\pm$ 0.      & 0.17   & 5        \\
  Terzan~9     & 270.41167  & -26.83972  & -1.36   &  68.49  $\pm$  0.56     &   5.77 $\pm$  0.34  &  2.46  $\pm$ 0.32   &   12.0    $\pm$ 1.4    & 0.295  & 23       \\
  Djorg~2  & 270.45438  & -27.82582  & -1.07   &  -149.75$\pm$  1.1      &   8.76 $\pm$  0.18  &  0.8   $\pm$ 0.13   &   12.5   $\pm$ 0.3   & 0.079  & 10       \\
  NGC~6517  & 270.46075  & -8.95878   & -1.58   &  -35.06 $\pm$  1.65    &   9.23 $\pm$  0.56  &  3.24  $\pm$ 0.26   &   19.5   $\pm$ 2.8    & 0.27   & 1        \\
  Terzan~10    & 270.74083  & -26.06694  & -1.62   &  211.37 $\pm$  2.27    &   10.21 $\pm$  0.40  &  2.17  $\pm$ 0.37   &   30.2   $\pm$ 5.6    & 0.191  & 2        \\
  NGC~6522  & 270.89198  & -30.03397  & -1.22   &  -15.23 $\pm$  0.49    &   7.30 $\pm$  0.21  &  1.04  $\pm$ 0.17   &   21.1   $\pm$ 1.3    & 0.181  & 15       \\
  NGC~6528  & 271.2067   & -30.05578  & -0.16   &  211.86 $\pm$  0.43    &   7.83 $\pm$  0.24  &  0.7   $\pm$ 0.1    &   5.7  $\pm$ 0.7   & 0.067  & 4        \\
  NGC~6539  & 271.20728  & -7.58586   & -0.74    &  35.19  $\pm$  0.5     &   8.17 $\pm$  0.39  &  3.09  $\pm$ 0.07   &   20.9   $\pm$ 1.7     & 0.317  & 1        \\
  NGC~6540  & 271.53566  & -27.76529  & -1.02 &  -16.5  $\pm$  0.78    &   5.91 $\pm$  0.27  &  2.34  $\pm$ 0.25   &   3.5  $\pm$ 1.2    & 0.165  & 6        \\
  NGC~6544  & 271.83383  & -24.99822  & -1.52   &  -38.46 $\pm$  0.67    &   2.58 $\pm$  0.06  &  5.62  $\pm$ 0.06   &   9.1  $\pm$ 0.6   & 1.078  & 27       \\
  NGC~6553  & 272.31532  & -25.90775  & -0.19    &  -0.27  $\pm$  0.34    &   5.33 $\pm$  0.13  &  2.83  $\pm$ 0.13   &   28.5    $\pm$ 1.6    & 0.494  & 17       \\
  NGC~6558  & 272.57397  & -31.76451  & -0.99   &  -195.12$\pm$  0.73    &   7.47 $\pm$  0.29  &  1.08  $\pm$ 0.17   &   2.65  $\pm$ 0.08  & 0.073  & 6        \\
  Terzan~12    & 273.06583  & -22.74194  & -0.56   &  95.61  $\pm$  1.21     &   5.17 $\pm$  0.38  &  3.17  $\pm$ 0.34   &   8.7  $\pm$ 2.0    & 0.312  & 6        \\
\hline
	\end{tabular}
\end{table*}

\setcounter{table}{0}
\begin{table*} 
\label{tab:list}
	\centering
	\caption{Continued}
	\begin{tabular}{lrrccrrccr} 
		\hline
		\hline
GC  &  $\alpha_{\rm cen}$ & $\delta_{\rm cen}$  &  $\langle{\rm \lceil Fe/H \rceil}\rangle$ &  $\langle{\rm R.V.}\rangle$ & $d_\odot$  &  $R_{\rm GC}$  &  Mass  &  $r_{\rm J} $ &  N\\
		\hline
 &  deg  & deg  &  & km~s$^{-1}$ & kpc & kpc & 10$^4$~M$_\odot$ & deg & \\
		\hline
  NGC~6569  & 273.41167  & -31.82689  & -0.92   &  -49.83 $\pm$  0.5     &  10.53 $\pm$  0.26  &  2.59  $\pm$ 0.23   &   23.6   $\pm$ 2.0    & 0.226  & 14       \\
  NGC~6642  & 277.97596  & -23.4756   & -1.09   &  -60.61 $\pm$  1.35    &   8.05 $\pm$  0.20  &  1.66  $\pm$ 0.01   &   3.4  $\pm$ 0.1   & 0.11   & 12       \\
  NGC~6656  & 279.09976  & -23.90475  & -1.70   &  -148.72$\pm$  0.78    &   3.30 $\pm$  0.04  &  5.0   $\pm$ 0.03   &   47.6   $\pm$ 0.5    & 1.308  & 412      \\
  NGC~6715  & 283.76385  & -30.47986  & -0.62   &  143.13 $\pm$  0.43    &  26.28 $\pm$  0.33  &  18.51 $\pm$ 0.32   &   178.0   $\pm$ 3.0    & 0.618  & 1809     \\
  NGC~6717  & 283.77518  & -22.70147  & -1.12   &  30.25  $\pm$  0.9     &   7.52 $\pm$  0.13  &  2.38  $\pm$ 0.02   &   3.6  $\pm$ 0.8    & 0.151  & 5        \\
  NGC~6723  & 284.88812  & -36.63225  & -1.02   &  -94.39 $\pm$  0.26    &   8.27 $\pm$  0.10  &  2.47  $\pm$ 0.02   &   17.7   $\pm$ 1.1    & 0.258  & 9        \\
  NGC~6752  & 287.7171   & -59.98455  & -1.47   &  -26.01 $\pm$  0.12    &   4.12 $\pm$  0.04  &  5.3   $\pm$ 0.02   &   27.6   $\pm$ 0.4   & 0.913  & 152      \\
  NGC~6760  & 287.80027  & 1.03047    & -0.75   &  -2.37  $\pm$  1.27    &   8.41 $\pm$  0.43  &  5.17  $\pm$ 0.14   &   26.9   $\pm$ 2.5    & 0.488  & 11       \\
  Palomar~10    & 289.50693  & 18.57899   & 0.02    &  -31.7  $\pm$  0.23     &   8.94 $\pm$  1.18  &  7.6   $\pm$ 0.59   &   16.2   $\pm$ 2.7    & 0.431  & 3        \\
  NGC~6809  & 294.99878  & -30.96475  & -1.76   &  174.7  $\pm$  0.17    &   5.35 $\pm$  0.05  &  4.01  $\pm$ 0.03   &   19.3   $\pm$ 0.8   & 0.549  & 98       \\
  NGC~6838  & 298.44373  & 18.77919   & -0.75   &  -22.72 $\pm$  0.2     &   4.00 $\pm$  0.05  &  6.86  $\pm$ 0.01   &   4.6  $\pm$ 0.2   & 0.619  & 129      \\
  NGC~7078  & 322.49304  & 12.167     & -2.29    &  -106.84$\pm$  0.3     &  10.71 $\pm$  0.10  &  10.76 $\pm$ 0.07   &   63.3   $\pm$ 0.7   & 0.757  & 155      \\
  NGC~7089  & 323.36258  & -0.82325   & -1.47   &  -3.78  $\pm$  0.3     &  11.69 $\pm$  0.11  &  10.54 $\pm$ 0.08   &   62.7   $\pm$ 1.1    & 0.548  & 36       \\		
\hline
	\end{tabular}
\end{table*}

\section{Selecting Cluster Member Candidates} \label{sec:selection}

The number of stars targeted per GC varied widely as a function of the number of visits, apparent magnitude, and apparent GC size, which constrains the number of possible targets in any given visit by virtue of the fiber collision radius of $\sim$~1'.  Moreover, the fraction of {\it bona fide} GC members from previous studies also varies widely across the GC sample, as does the number of targets lacking a previous membership assignment based on radial velocity or proper motion measurement.

\begin{table}
\centering
\caption{Definition of two subgroups of candidate members. Column information: (1) Subgroup type; (2) angular distance limits in units of the Jacobi radius, given in Table~\ref{tab:list}; (3) PM limits; (4) RV limits, 
(5) $\lceil {\rm Fe/H} \rceil$ limits.  
Limits in columns (3) to (5) in units of residuals defined as 
$\delta~X = \frac{{\rm res}(X)}{\sigma_X}$, 
where ${\rm res}(X) = | X - \mu_X | $, and $\mu_X$ 
and $\sigma_X$
stand for the mean and r.m.s. scatter of each observable, respectively.
}
\begin{tabular}{lcccc} 
\hline
Subgroup & Distance & PM & RV & $\lceil {\rm Fe/H} \rceil$ \\
(1) & (2) & (3) & (4) & (5) \\
\hline
Likely & $r < r_{\rm J}$ & $\delta~{\rm PM} <2$ & $\delta~{\rm RV} <2$ & any \\
Outlier & $r_{\rm J} < r < 4~r_{\rm J}^\star$ & $2 < \delta~{\rm PM} <10$ & $2 < \delta~{\rm RV} <3$ & $\delta~\lceil{\rm Fe/H}\rceil <2$\\
		\hline
\end{tabular}
\begin{tablenotes}
\item $^\star$  For GCs located in crowded fields towards the inner Galaxy, the initial search radius was reduced from 4~\rJ\ to 2~\rJ, in order to minimise contamination by field stars.
\end{tablenotes}
\label{tab:subgroups}
\end{table}

\begin{figure}
	\includegraphics[width=\columnwidth]{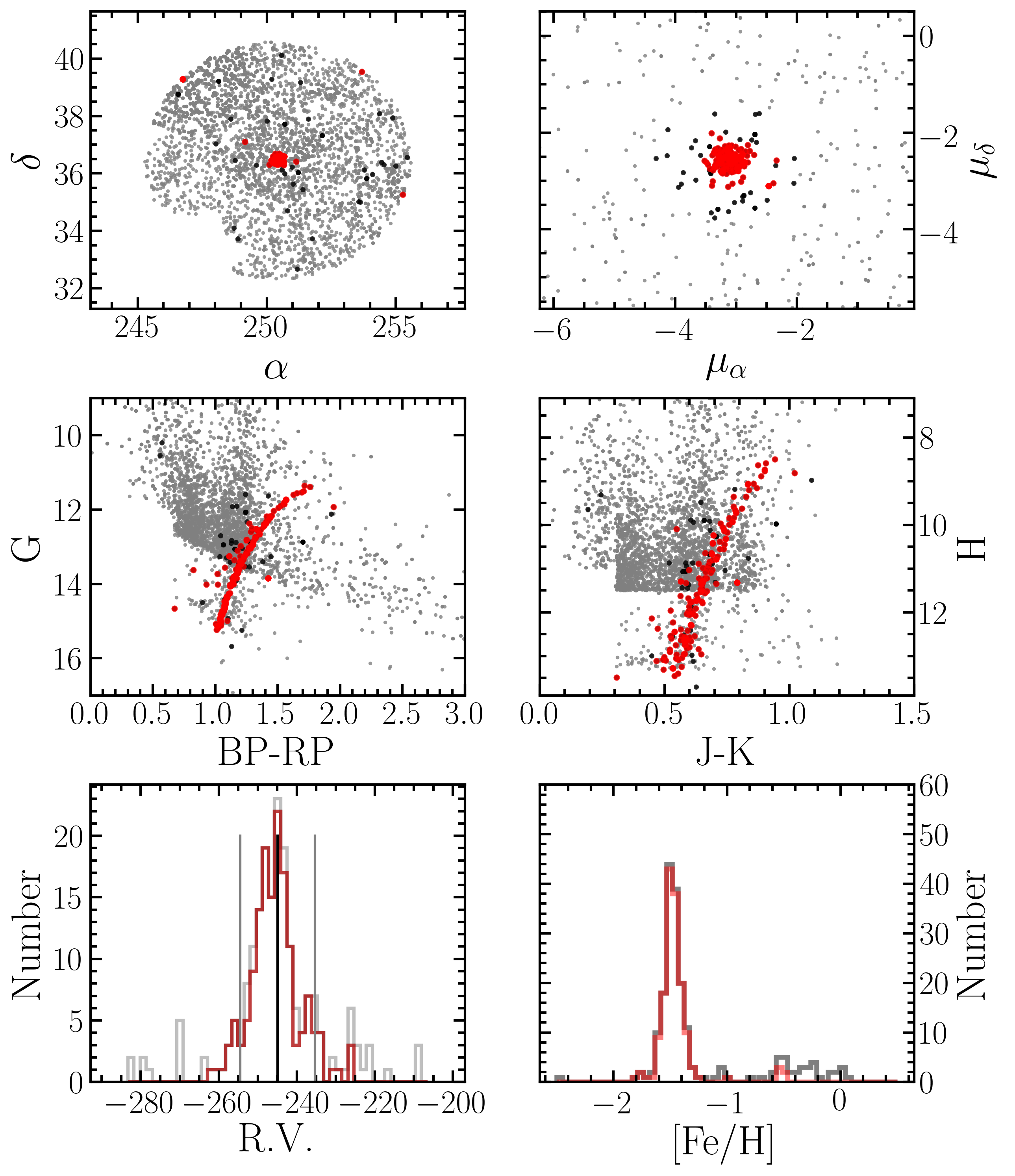}
    \caption{Selecting M13 members in the APOGEE catalogue.  In all panels, grey dots represent catalogue stars within 4~r$_J$ of the GC centre.  Black dots represent stars whose proper motions differ from the mean value from the VB catalogue by no more than 4~$\sigma_{\rm PV}$.  Red dots represent a sub-sample of the former whose radial velocities differ from the GC mean by no more than 3~$\sigma_{\rm RV}$.}
    \label{fig:selection_good}
\end{figure}

\begin{figure}
	\includegraphics[scale=0.3]{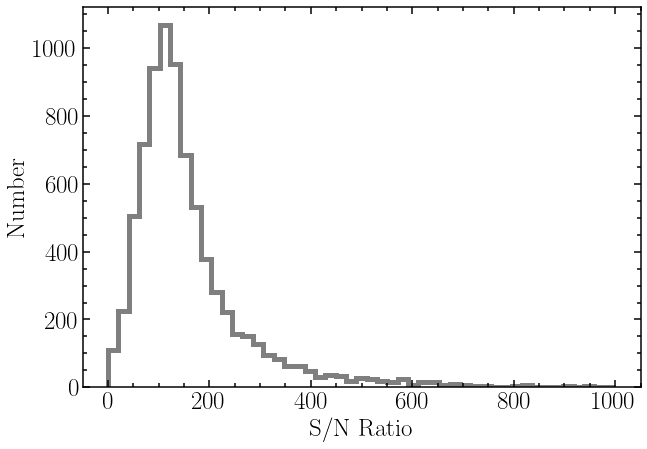}
    \caption{Distribution of the S/N/pixel of the resulting sample.}
    \label{fig:sn_hist}
\end{figure}

The situation mandates a strategy based on a sweeping search of the entire APOGEE DR17 catalogue for GC members defined according to a set of homogeneous membership criteria.  By proceeding in this way we hope to generate a catalogue that confirms previously established memberships while further extending  member samples on the basis of good quality radial velocities and proper motions.

The philosophy underlying our approach is to generously consider every star with a reasonable probability of belonging to a given GC, providing elements to enable the catalogue users to make their own informed sample selections.  In short, catalogue completeness is prioritised over purity.  Nevertheless, the catalogue is devised in such a way as to make a conservative selection leading up to a very pure sample quite straightforward.

Stars are selected on the basis of angular distance from GC centre, proper motion (PM), and radial velocity (RV) only.  Criteria for selection are defined in terms of the GC's Jacobi radius ($r_{\rm J}$), as well as central values and dispersion of PMs ($\mu_{\rm PM}$ and $\sigma_{\rm PM}$) and RVs ($\mu_{\rm RV}$ and $\sigma_{\rm RV}$).  We decided not to use position in the colour-magnitude diagram as a selection criterion, to avoid biasing against possible minority populations. We adopt the following sets of criteria to define two broad types of candidate members:

\begin{itemize}
    \item {\it Likely members} are those meeting a {\it strict} set of angular distance, PM, and RV criteria, {\it regardless of their chemical compositions}.
    \item {\it Outliers} are stars meeting more relaxed angular distance, PM, and RV criteria, whose metallicities match closely those of nearby GCs.
\end{itemize}

The {\it Likely} group, as its name indicates, contains the stars that have the highest probability of being cluster members.  By not imposing a metallicity condition to define this group we wish to avoid missing members for which ASPCAP could not find a metallicity solution (the case of very warm stars) or those with potentially large errors in metallicity, (the case of both very warm and very cool stars or those with low S/N spectra).  Moreover, {\it Likely} members with very discrepant metallicities could represent a fringe GC population.  Conversely, the {\it Outlier} group contains stars whose chemical compositions are consistent with membership, but whose position and kinematics suggest at best a loose association.  Inclusion of the {\it Outlier} group aims at catching a maximum number of extra-tidal stars.  In cases of GCs presenting a large spread in metallicity, such as $\omega$~Cen, M~54, Terzan~5, or Liller~1, similarity in terms of [Fe/H] cannot be used to define {\it Outliers}, although those with abundance patterns consistent with a second-generation nature are retained and flagged.  

The quantitative definitions of these two sub-groups is provided in Table~\ref{tab:subgroups}.  GC centre coordinates, as well as the values for $r_{\rm J}$, $\mu_{\rm PM}$, $\mu_{\rm RV}$, and $\sigma_{\rm RV}$ were adopted from the VB catalogue.  The generous upper angular distance limits adopted for {\it Outliers} is aimed at enabling the identification of extra-tidal GC members.  The method to estimate $\sigma_{\rm PM}$ is described below.  The very generous PM threshold was adopted after we found out that some good candidates were located several $\sigma_{\rm PM}$ off of the mean PM value, which may reflect our admittedly rough estimated of $\sigma_{\rm PM}$.  

Our procedure can be summarised as follows.  We start by obtaining an estimate of the proper motion dispersion, $\sigma_{\rm PM}$. Data from the {\it Gaia} eDR3 archive were downloaded for each cluster. Adopting mean proper-motion values from the VB catalogue we calculated $\sigma_{\rm PM}$ through a single $\sigma$-clipping iteration aimed at removing background contamination.  That measured dispersion is obviously larger than the intrinsic dispersion, since it folds in measurement errors which are not the same for every cluster.  Given those estimates, stars are considered to be {\it Likely} GC members if they meet the set of strict criteria listed on the first row of Table~\ref{tab:subgroups}. Next the APOGEE catalogue was searched for {\it Outliers}, by following the set of loose criteria listed on the second row of Table~\ref{tab:subgroups}.  The selection process is illustrated in Figure~\ref{fig:selection_good}.

\begin{figure*}
	\includegraphics[width=\textwidth]{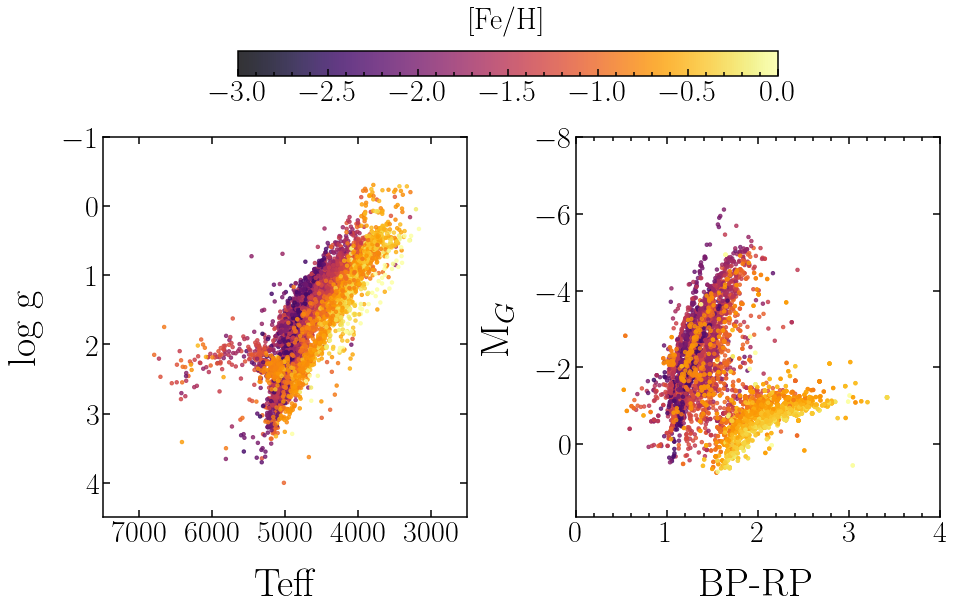}
    \caption{{\it Right:} Kiel Diagram for the resulting sample. Note that this plot does not display all stars included in the catalog, as ASPCAP failed to deliver stellar parameters for some stars. {\it Left: Gaia} eDR3 CMD of the GC parent sample, including only stars with A$_{\rm K}<0.3$. In both panels stars are colour-coded by the APOGEE DR17 Fe abundances.}
    \label{fig:Kiel_CMD}
\end{figure*}

The resulting sample consists of a total of 7,737 entries for 6,424 unique candidate members associated with 72 GCs.  Multiple entries occur for a number of stars located in overlapping fields and/or observed as part of different programs.  The quality of the data is illustrated in Figure~\ref{fig:sn_hist}, which shows the distribution of the median S/N/pixel of the resulting sample, where $\sim$93\% of the spectra have S/N$>$50.  The distributions of the stars in the Kiel diagram and Gaia colour-magnitude diagram are shown in \ref{fig:Kiel_CMD}.  The right panel of shows the distribution of the sample stars in the {\it Gaia} undereddened colour-magnitude diagram, where the range of GC metallicities can be immediately appreciated.  In the left panel, sample stars are displayed in the Kiel diagram, which brings to sharp relief the high precision of APOGEE stellar parameters.


\section{Membership Probabilities} \label{sec:probs}

In order to provide users of this catalogue with the elements required for deciding which samples should be considered for their analysis, two sets of membership probabilities are provided.  The first set is based on a Gaussian mixture modelling of the {\it Gaia} eDR3 positions and proper motions of GC stars, and directly imported from the VB catalogue (Section~\ref{sec:VBprobs}).  In addition, we derive our own set of independent membership probabilities, based on the APOGEE radial velocities. 

\subsection{Vasiliev \& Baumgardt probabilities} \label{sec:VBprobs}



For the user's convenience we briefly summarise the membership probability estimates provided in the VB catalogue.  For further details the user is referred to the original papers \citep{Vasiliev2021,Baumgardt2021}.  Membership probabilities were determined via a mixture modelling approach from which they also infer cluster properties such as mean parallax, proper motion, dispersion and structural parameters.  The initial sample is obtained by extracting all sources with 5- or 6-parameter astrometric solutions from {\it Gaia} within a certain distance from the centre of each GC, which in the general case is taken to be a few times greater than the cluster half-light radius. A first run of mixture modelling in the 3D astrometric space is performed on a subset of the sources with the most reliable astrometry, where one of the Gaussian components represents the cluster and the remaining component(s) account for the field stars.  A full mixture model is then run, where a Plummer model is adopted to match each GC's density profile, with the scale radius as a free parameter.  The parameter space is explored with a Markov Chain Monte Carlo code initialised with astrometric parameters determined by extreme deconvolution.  Membership probabilities for each star are then determined following convergence of the MCMC runs.  Colour-magnitude diagrams of members thus obtained for each GC are inspected visually to verify the outcome of the mixture model, which did not utilise any the photometric information. Finally, the mean parallax and proper motion of each cluster and their uncertainties are taken from the MCMC chain.

\bigskip

\subsection{RV-based  Probabilities} \label{sec:rvprobs}

Exceedingly accurate radial velocities are one of the main data products of the APOGEE survey.  This can be verified through a quick comparison with the data from latest Gaia release \citep[DR3,][]{gaiadr3}.  We cross-matched our sample for M~13 with the Gaia DR3 catalogue, obtaining 108 matches.  The mean heliocentric radial velocity and r.m.s. scatter for each sample are in excellent agreement, with ${\rm <rv_{APO}>}=-246.30~{\rm km~s}^{-1}$ and ${\rm <\sigma_{APO}>}=5.27~{\rm km~s}^{-1}$ for the APOGEE sample, and ${\rm <rv_{Gaia}>}=-246.33~{\rm km~s}^{-1}$ and ${\rm <\sigma_{Gaia}>}=5.95~{\rm km~s}^{-1}$.  It is noteworthy that the mean radial velocities agree to within 30~${\rm m~s^{-1}}$, reflecting the great accuracy of the two data sets. In addition, the r.m.s. scatter, which results from the convolution between the cluster velocity dispersion and measurement error, is lower in the APOGEE sample by $\sim$12\%, reflecting APOGEE's superior radial velocity precision.

We take advantage of this high-quality data set to complement the membership information available from the VB catalogue with RV based membership probabilities.  These probabilities were estimated as follows.  We adopted a procedure similar to that of \cite{Gieseking1985} whereby the RV distribution within the field of each GC was modelled as a combination of Gaussian functions plus a constant background.  For any given star $i$, the RV-based membership probability is given by:
\begin{equation}
p_i = {\rm \frac{G_{\rm gc}(v_i)}{G_{\rm gc}(v_i) + B(v_i)}}
\end{equation}
Where ${\rm v_i}$ is the radial velocity of the star, ${\rm G_{gc}(v)}$ is the Gaussian function describing the RV distribution of the GC, and ${\rm B(v)}$ is a function accounting for the RV distribution of the field background. 

\begin{figure}
	\includegraphics[width=\columnwidth]{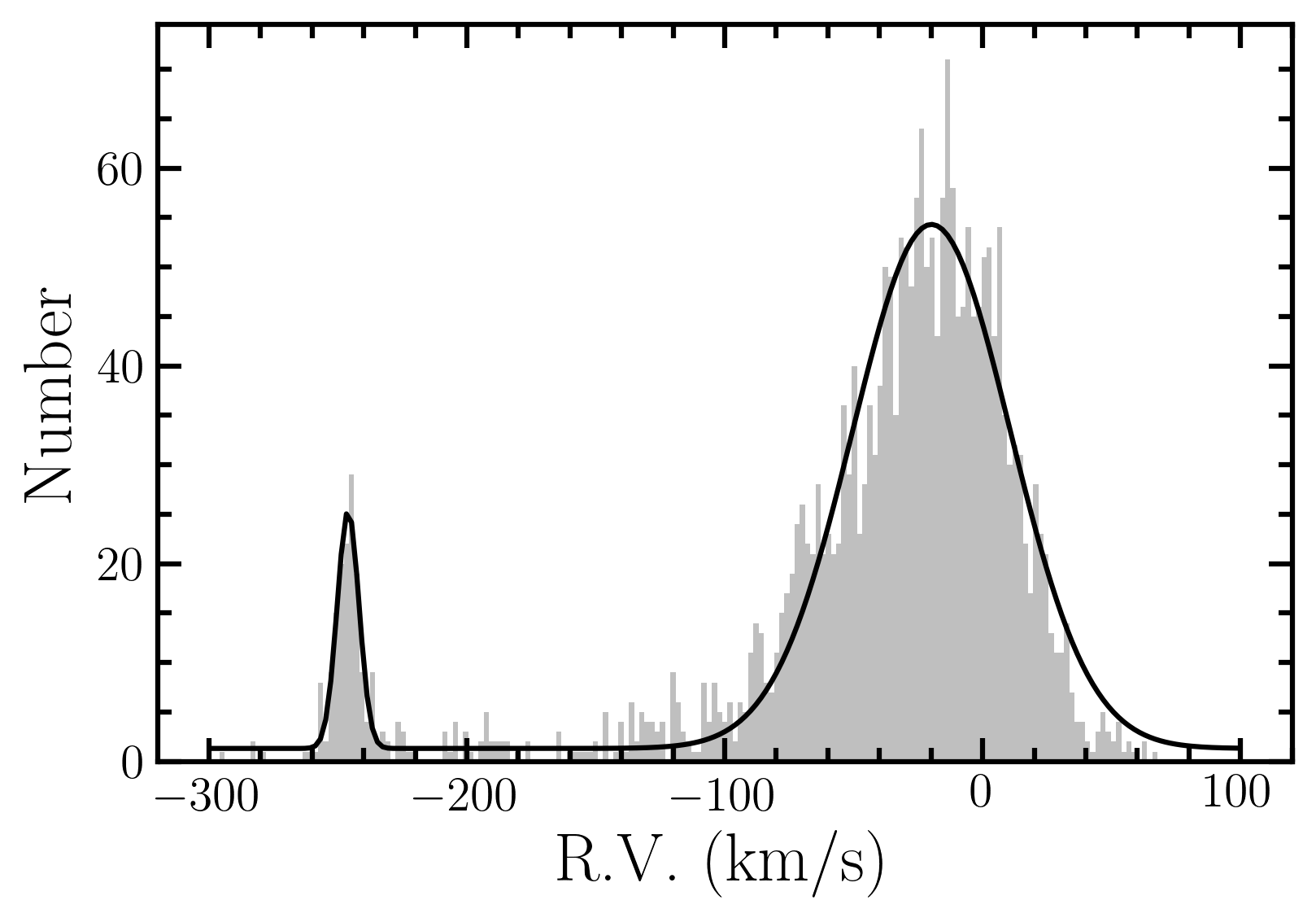}
    \caption{Fit of the RV distribution in the field of M13.  The model fit is a double Gaussian with an additional constant background.  The secondary peak corresponds to the RVs of the cluster.}
    \label{fig:RVs}
\end{figure}

For well-sampled GCs, the functions ${\rm G_{gc}(v)}$ and ${\rm B(v)}$ were obtained from a fit to the RV distribution from the stars contained within the field of each GC.  In cases where the GC is poorly sampled and/or the contrast with the background is poor, the ${\rm G_{gc}(v)}$ function adopted was based on parameters (mean RV and velocity dispersion) gathered from the VB catalogue.  The background function ${\rm B(v)}$, in the general case, was a combination of Gaussians and a constant floor value.  In no case were more than two Gaussians required to account for the background data.  An example fit is shown in Figure~\ref{fig:RVs}.

\begin{figure*}
	\includegraphics[width=\textwidth]{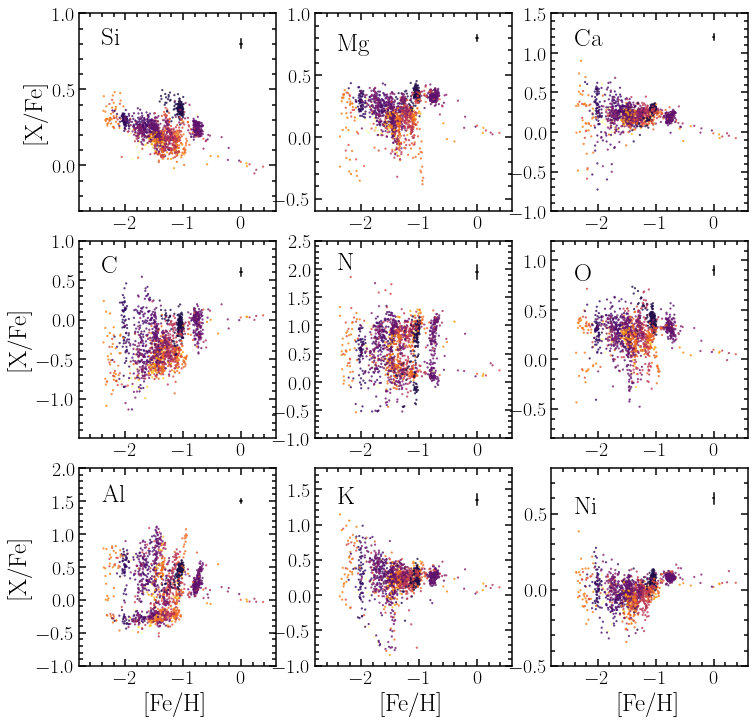}
    \caption{Sample elemental abundances for Galactic GCs included in the VAC. Only abundances derived from spectra with S/N$>$150 are shown.  To distinguish individual GCs, data are colour-coded by the cluster heliocentric distance.  GCs with large spreads in metallicity, namely $\omega$~Cen and M~54, are excluded from this plot.  Mean error bars are displayed on the top right of each panel.}
    \label{fig:abunds}
\end{figure*}

\section{Results and Science Highlights} \label{sec:results}

This value-added catalogue can be employed in a myriad of different science projects.  We highlight a few aspects of the data base that illustrate its potential.  In Figure~\ref{fig:abunds}, selected elemental abundances sampling different nucleosynthetic pathways are displayed in various panels.  Only abundances derived from spectra with S/N$>$150 are shown.  To distinguish stars associated with individual GCs,  symbols are colour-coded by heliocentric distance.  The complexity of the GC member candidates distribution in chemical-composition space is promptly evident from a first glance to these data.

\begin{figure}
	\includegraphics[width=\columnwidth]{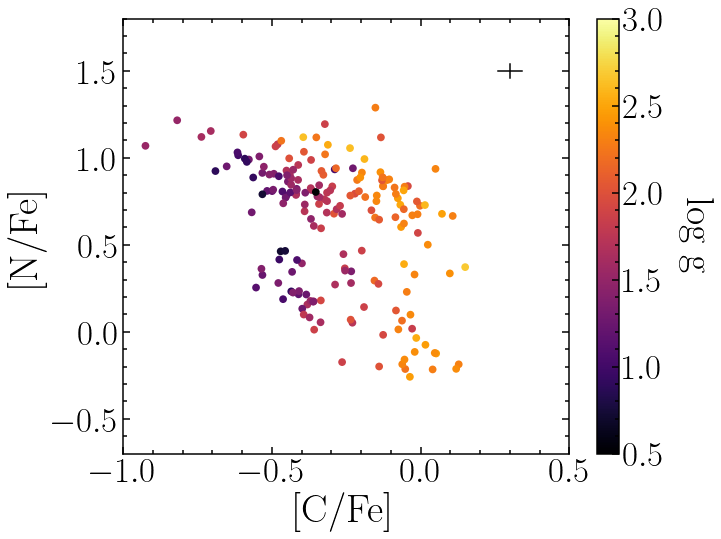}
    \caption{Carbon-Nitrogen anti-correlation in the globular cluster M5 (NGC~5904).  Symbols are colour-coded by surface gravity ($\log g$) to distinguish C-N abundance variations due to stellar evolution from those associated with the multiple populations phenomenon.  Two diagonal sequences can be seen.  Along each sequence, the variations of N and C abundances are correlated with $\log g$, as deep mixing brings the byproducts of the CNO-cycle to the star's surface steadily changing its chemical composition during evolution along the giant branch.  The more drastic anti-correlation due to the MP phenomenon connects stars with same $\log g$ between the two sequences. Mean error bars are displayed on the top right.}
    \label{fig:NC}
\end{figure}

In Figure~\ref{fig:NC}, the data for M~5 (NGC~5904) are displayed on the [C/Fe] vs [N/Fe] plane, where symbols are colour-coded by surface gravity ($\log g$).  Two sequences are clearly visible, where a gentle variation of N and C abundances can be seen to be correlated with $\log g$. This variation is due to mixing along the giant branch, whereby more evolved stars (lower $\log g$) display depleted C and enhanced N due to the progressive mixing of CNO-processed material during the evolution along the red giant branch.  The more drastic variation associated with the MP phenomenon connects stars with same $\log g$ between the two sequences \citep[e.g.,][]{Phillips2022}.

\begin{figure}
\includegraphics[width=\columnwidth]{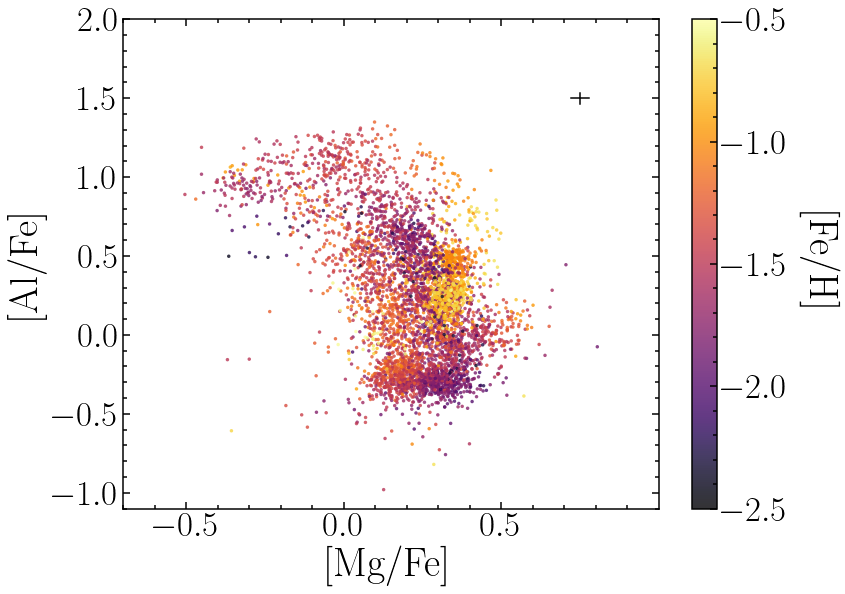}
    \caption{Magnesium-Aluminium anti-correlation for a collection of GCs with [Fe/H]$\leq-0.5$.  Different sequences are displaced on this plane according to GC natal chemical composition. The Mg-Al anti-correlation is weakened or even absent towards higher metallicity. Mean error bars are displayed on the top right.}
    \label{fig:AlMg}
\end{figure}

In Figure~\ref{fig:AlMg}, data for various GCs with [Fe/H]$<-0.5$ are displayed on the [Mg/Fe] vs [Al/Fe] plane.  Symbols are colour-coded by metallicity.  Metal-poor GCs show a strong anti-correlation between these two elements.  The various GC sequences are displaced relative to each other due to variations in the systems' natal chemical compositions, associated with their origin.  The weakening of this anti-correlation with increasing metallicity \citep[e.g.,][]{Nataf2019} manifests itself by the near absence of an anti-correlation in the most metal-rich GCs.

Finally, in the Appendix we present a comparison of the APOGEE DR17 elemental abundances published in this value added catalogue with data from various sources from the literature.

\subsection{Extra-tidal candidates} \label{sec:ets}

Globular clusters are slowly dissolving, shedding stars under the combined effect of evaporation and tidal stripping as they follow their orbits within the Milky Way dark matter halo.  Evidence to this phenomenon has been documented as stars are detected beyond GC tidal radii, in the form of tidal streams \citep[e.g.,][]{Odenkirchen2001,Belokurov2006,Grillmair2006,BonacaHogg2018,Malhan2021} and diffuse outer envelopes or less defined collections of extra-tidal stars \citep[e.g.,][]{Kuzma2016,Kuzma2018,Chun2020,Kundu2022,Piatti2022}. 

Identifying extra-tidal stars is a difficult task requiring deep photometry over a wide field of view.  More recently, data from the Gaia satellite have enabled the use of proper motions for that purpose \citep[e.g.,][]{Kundu_2019}. In the past decade, chemical tagging has been used to identify field stars with chemistry that is characteristic of GC populations \citep[e.g.,][]{Martell2010,Lind2015,Martell2016,S2017,Trincado2017,Tang2019}, leading up to moderately robust estimates of the contribution of dissolved GCs to the Milky Way stellar halo mass budget \citep[e.g.,][]{Martell2011,S2017,Koch2019,Horta2021}.  

Linking so-called ``N-rich'' field stars with their parent GCs is quite important as a means to establish once and for all their GC origin \citep[e.g.,][]{Kisku2021}.  However, such associations have proved  difficult, resulting from likelihood estimates based on orbital parameters \citep[e.g.,][]{SavinoPosti2019}.  

Detailed chemistry and precision radial velocities for large samples, combined with {\it Gaia}-quality astrometry and GC structural parameters can make an important contribution in this context.  Large samples with precision chemistry enables unequivocal association of extra-tidal stars with their parent GCs.  Indeed, recent work has provided evidence for the presence of N-rich stars beyond the Jacobi radius  of M~54 and Palomar~5 \citep{Trincado2021,Phillips2022}.  

\begin{figure}
	\includegraphics[width=\columnwidth]{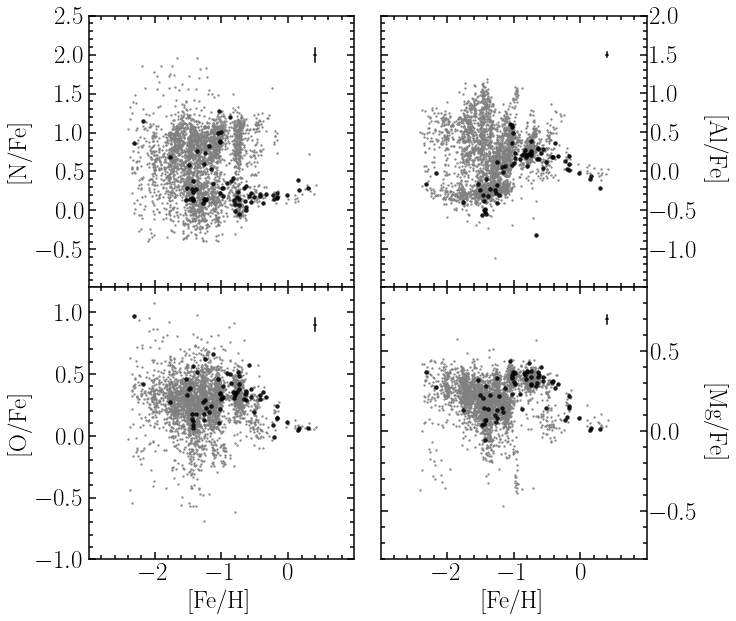}
        \caption{Candidate GC members in various chemical planes.  Stars within their parent GC Jacobi radii are shown as grey symbols whereas extra-tidal stars are displayed as black dots.  A substantial fraction of the extra-tidal stars have N-rich abundance patterns, confirming the GC-origin of N-rich stars identified in previous studies. For previous identifications of extra-tidal N-rich stars see discussion in text. Mean error bars are displayed on the top right of each panel.}  
    \label{fig:ets}
\end{figure}

In Figure~\ref{fig:ets} VAC data are displayed on various chemical planes.  Data for the $\omega$~Cen and M~54 are omitted from these plots.  Grey dots show the whole sample, and black dots represent only stars located beyond the Jacobi radius of their parent GC.  While most extra-tidal stars have normal chemistry, a few dozen N-rich stars can be identified in those planes, due to their enhanced abundances of N and Al, and depleted Mg and O.  Extra-tidal stars can be easily identified in the VAC by the value of the parameter {\tt DPOS}, which is equal to the angular distance to each GC centre, in units of $r_{J}$.  Extra-tidal stars have {\tt DPOS}~>~1.

\subsubsection{The case of M~54}

\begin{figure}	\includegraphics[width=\columnwidth]{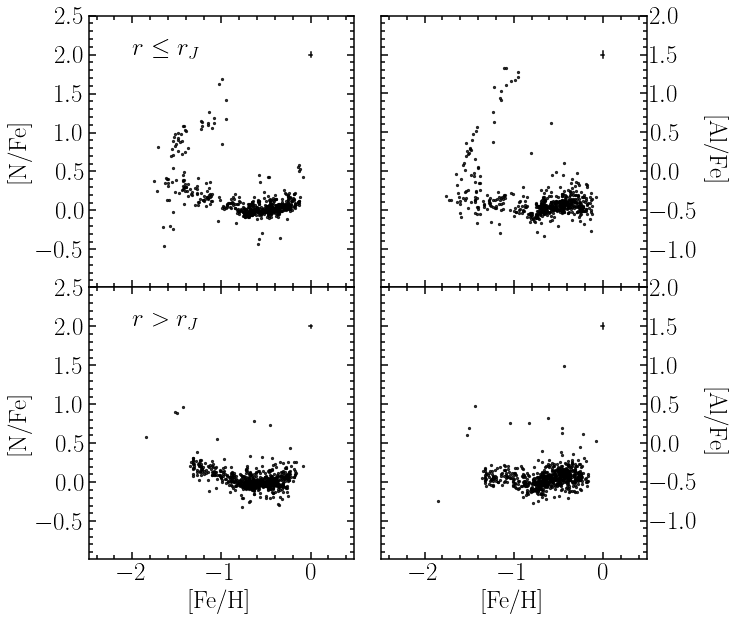}
    \caption{Candidate members of M~54  (NGC~6715), the nuclear cluster of the Sagittarius dwarf spheroidal, on the N-Fe and Al-Fe chemical planes.  Stars located within (top panel) or beyond (bottom) the cluster's Jacoby radius are displayed.  Note the large number of extra-tidal N-rich and/or Al-rich stars (bottom panel).  It is not clear whether this effect is real or due to an underestimate of M~54's Jacobi radius. Mean error bars are displayed on the top right of each panel.}
    \label{fig:M54}
\end{figure}

M~54 is the nuclear cluster of the Sagittarius dwarf Spheroidal (Sgr~dSph).  Its chemodynamical properties have been studied extensively \citep[e.g.,][]{Law_2010,Mucciarelli_2017} and merit some attention.  In Figure~\ref{fig:M54} we show the data for M~54 members on the same chemical planes as Figure~\ref{fig:ets}.  Top/bottom panels show intra/extra-tidal stars.  It is noteworthy that this cluster is characterised by a very large population of extra-tidal stars, some of which have N-rich abundance patterns \citep[see, e.g.,][]{Trincado2021}.  Indeed, a large fraction of the entire population of extra-tidal stars identified in this work are associated with M~54.  That could be a result of the cluster's undergoing severe tidal disruption under the MW potential, or rather reflect a possible underestimate of the M~54's Jacobi radius.  Such estimates are plagued by considerable uncertainties.  In the case of M~54, the situation is made worse by the fact that it is not known whether the cluster is positioned at the centre of its host galaxy's potential well, and whether it possesses its own dark matter halo \citep[e.g.,][]{Carlberg2022}.  In view of these uncertainties, we decide to retain a large number of M~54 candidate members, while acknowledging the reality that this sample is considerably contaminated by Sgr~dSph field stars.  The catalogue users are again provided with data they can use to select sub-samples according to their science goals.



\subsection{Abundance spreads and global parameters}
\label{sec:alspreads}

As discussed in the Introduction, perhaps the most puzzling observational feature of GCs is the presence of large anti-correlated spreads of the abundances of light elements.  Despite many efforts from various groups, no particular scenario has been able to account for this phenomenon in a quantitative fashion \citep[see review by][]{BastianLardo2018}.  
Naturally, correlations between chemical-composition spreads and GC global parameters can provide valuable constraints on formation models.  In this Section we provide a brief foray into the topic, exploring how this new catalogue can potentially contribute to this discussion.  We focus on Al spreads.  
Aluminium abundances are exceptionally well-measured in APOGEE spectra, over a wide range of metallicities.  Moreover, unlike nitrogen, aluminium spreads can be assessed in a fairly unambiguous way, since the abundance of this element is not affected by stellar evolution effects.

\begin{figure}	\includegraphics[width=\columnwidth]{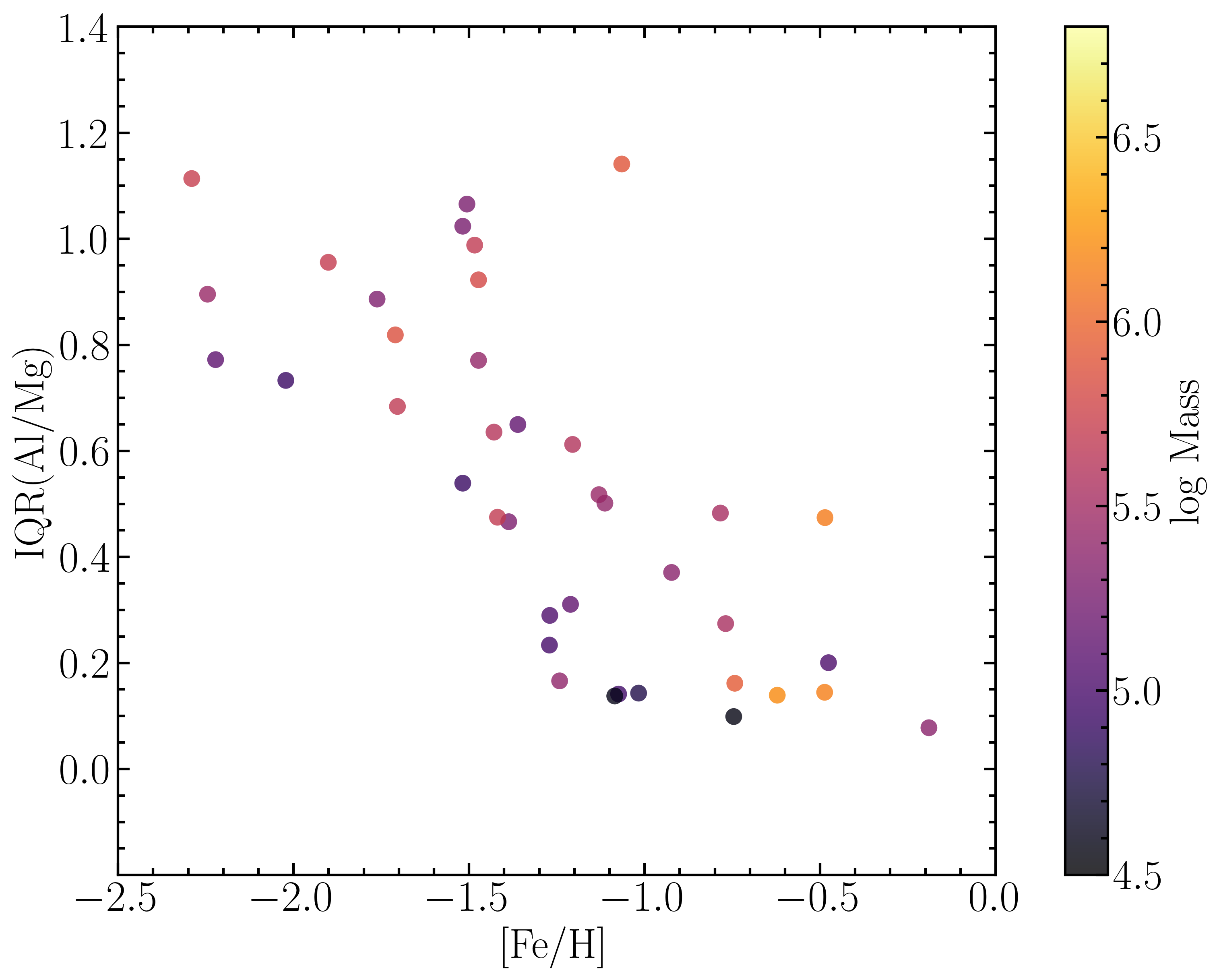}
    \caption{Inter-quartile range of the [Al/Mg] ratio plotted against GC metallicity.  A remarkable anti-correlation between the two quantities is apparent, with high significance ($\rho_x = -0.76$).  Symbols are colour-coded by log GC mass, but no correlation is apparent.}
    \label{fig:iqr_feh}
\end{figure}

Following \cite{Carretta2010}, who adopted the [O/Na] inter-quartile range as a measure of abundance spreads, we measure the inter-quartile range of the [Al/Mg] ratio.  We first examine the well known anti-correlation of aluminium spreads with GC metallicity \citep[see also][]{Nataf2019,Meszaros2020}.  
The data are displayed in Figure~\ref{fig:iqr_feh}, where a very clear anti-correlation between IQR(Al/Mg) and [Fe/H] is present, with a Spearman Rank correlation coefficient $\rho_x=-0.76$.  This result confirms previous studies reporting a substantial decrease of Al spreads in high-metallicity GCs.  Symbols are colour-coded by GC mass, but no clear correlation with that parameter can be seen.

\begin{figure}	\includegraphics[width=\columnwidth]{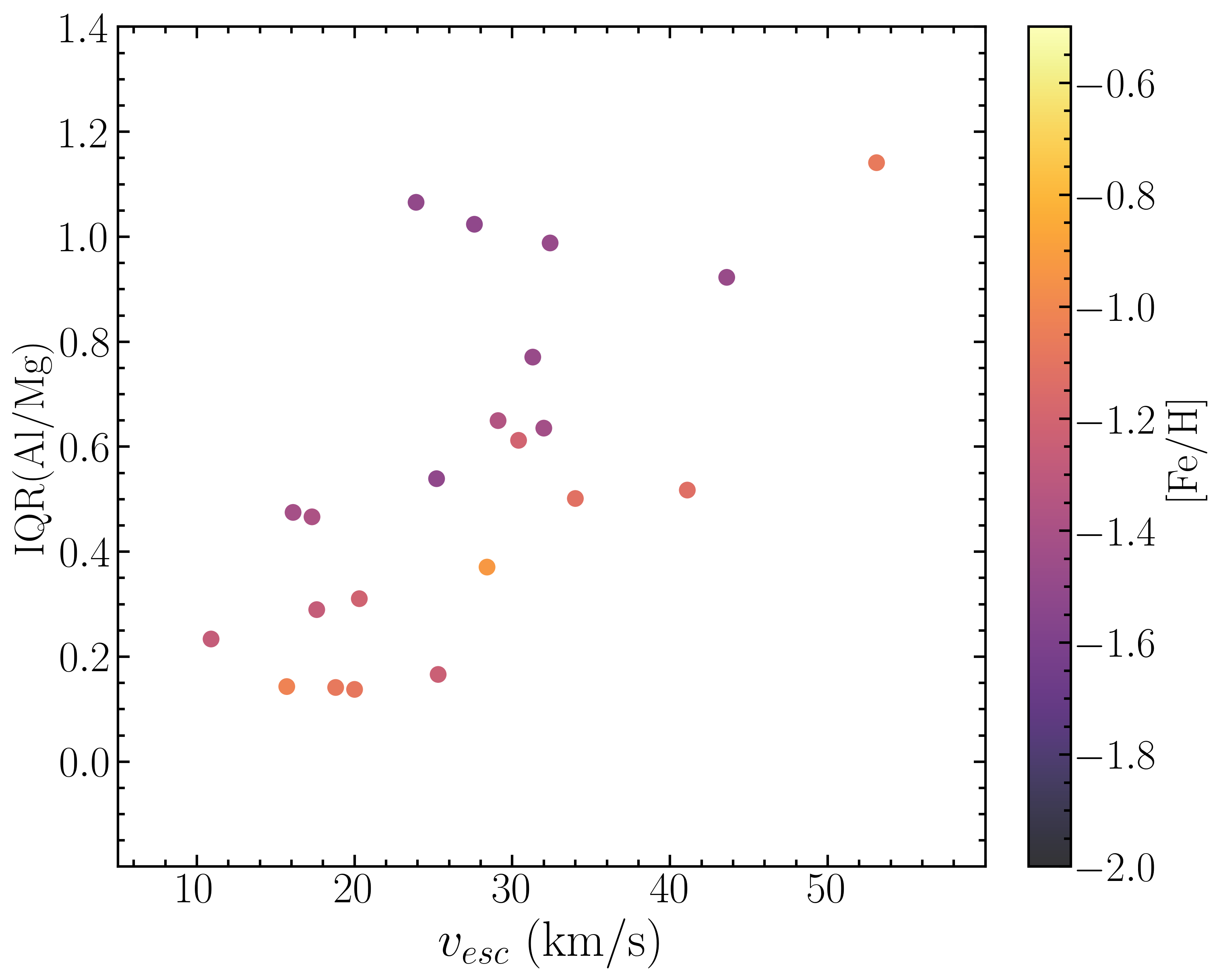}
    \caption{Inter-quartile range of the [Al/Mg] ratio plotted against central escape velocity.  A strong correlation is present ($\rho_x=0.68)$, but only after controlling for the effect of metallicity.  Only GCs with --1.7<[Fe/H]<--0.8 are considered.  Symbols are colour-coded by [Fe/H].
    This might be an indicator of chemical enrichment due to a history of feedback-regulated star formation.}
       \label{fig:iqr_vesc}. 
\end{figure}

Next, we examine the presence of a correlation between abundance spreads and a quantity related to a GC's gravitational potential.  Such a correlation is interesting, as it may be an indication of the presence of chemical enrichment brought about by a history of feedback-regulated star formation \cite[see also][]{Carretta2010,Schiavon2013,Sakari2016}.  
In Figure~\ref{fig:iqr_vesc}, we plot IQR(Al/Mg) against central escape velocity, from the VB catalogue.  Because the correlation between IQR(Al/Mg) and metallicity is so strong, we must control for this parameter, so only GCs with --1.7<[Fe/H]<--0.8 are shown.  
A strong correlation is seen ($\rho_x=0.68$).  We also find a strong correlation with central velocity dispersion ($\rho_x=0.69$) and GC mass ($\rho_x=0.62$).

\begin{figure}	\includegraphics[width=\columnwidth]{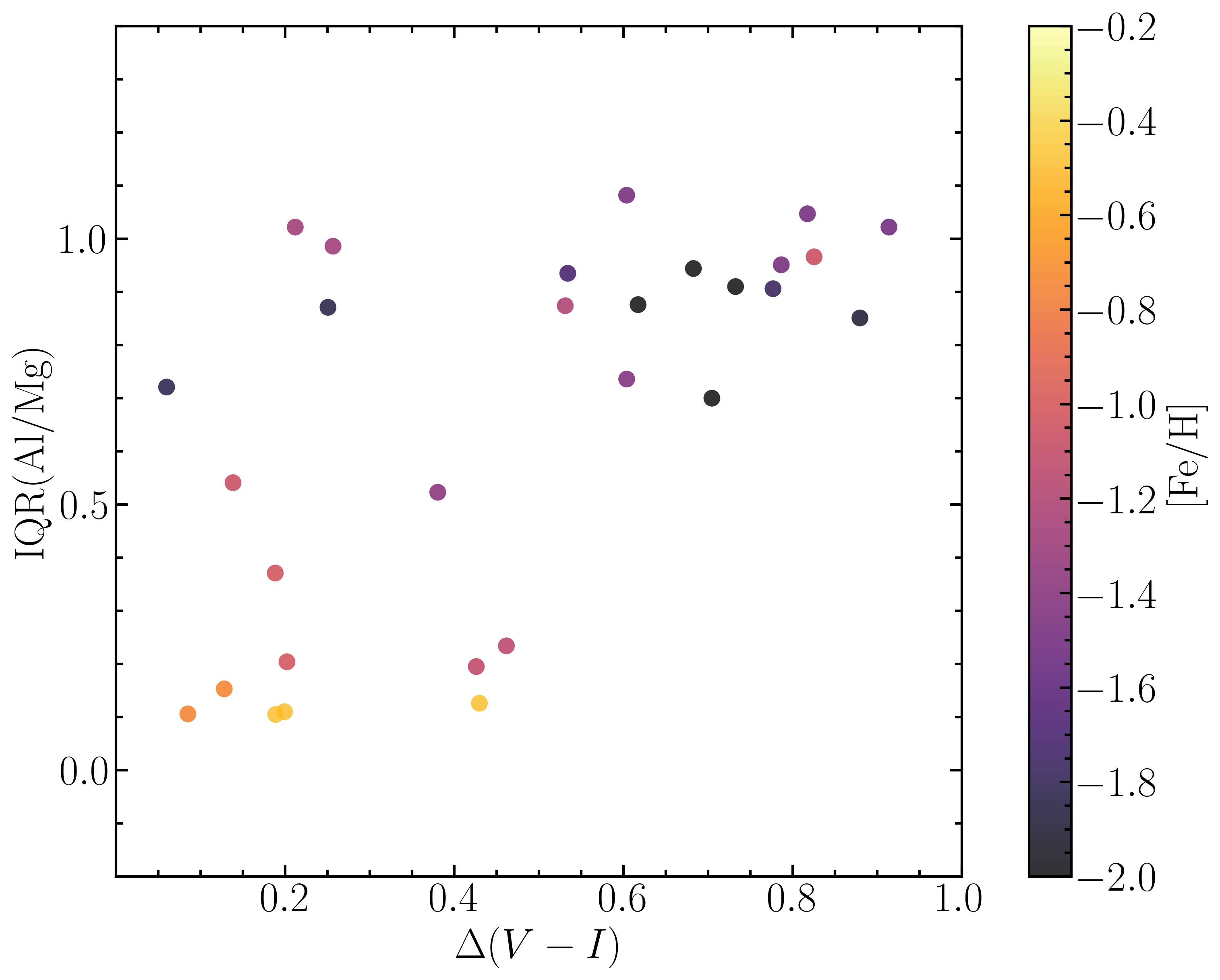}
    \caption{Inter-quartile range of the [Al/Mg] ratio plotted against a horizontal-branch morphology parameter, $\Delta$(V--I), which is higher for bluer horizontal-branch morphologies.  A fairly strong correlation is present ($\rho_x=0.63)$, in the sense that only GCs with high IQR(Al/Mg) present a blue HB.  Although GCs with red HBs can also have high IQR(Al/Mg), all GCs with low IQR(Al/Mg) have red HBs.}
    \label{fig:iqr_dvi}.
\end{figure}

We conclude by inspecting the relation between abundance spread and horizontal-branch (HB) morphology.  A correlation between these observatbles is expected because the morphology of the HB is in part dictated by the abundance of helium, an element for which there is strong evidence for abundance spreads \citep[e.g.,][]{Renzini2008}.  In the following, we adopt IQR(Al/Mg) as a surrogate for a spread in the abundance of helium.
The data are displayed in Figure~\ref{fig:iqr_dvi}, where IQR(Al/Mg) is plotted against the $\Delta$(V--I) parameter from \cite{Dotter2010}.  High values of $\Delta$(V--I) correspond to blue HB morphology.  
Although we find a relatively high Spearman rank correlation coefficient ($\rho_x=0.63$), the data behave in a subtle way.  There is a zone of avoidance at low IQR(Al/Fe) and blue HB morphology.  GCs with large abundance spreads can have either a red or a blue HB, but those with low spreads are all characterised by a red HB.  This may be related to the fact that the morphology of the horizontal branch is affected by a number of parameters besides He abundance, including age, binarity, and mass loss during the first-ascent red giant-branch phase.

\section{The Catalogue} \label{sec:summary}

The value-added catalogue presented in this paper consists of two files in FITS format.  The catalogue itself is contained in file {\tt VAC\_GC\_DR17\_synspec\_rev1.fits}, which includes all the data from the APOGEE DR17 {\tt allStar-dr17-synspec\_rev1.fits} for each of the 7,737 entries associated with GC candidate members.  
This file also incorporates distances from GC centres (in units of $r_{\rm J}$), residual proper motions, radial velocities, and [Fe/H], in units of the r.m.s. dispersions of those values.  Two sets of membership probabilities are also provided, those based on the radial velocity analysis in Section~\ref{sec:rvprobs} and those from the VB catalogue, when available.  
Another file, named {\tt GC\_parameters\_VAC.fits} contains, for each GC, the mean and r.m.s. values for RVs, proper motions, and metallicities, as well as a number of global parameters from the literature.  Both files are available for download from the SDSS DR17 value added catalog webpage  ({\tt https://www.sdss4.org/dr17/data\_access/value-added\-catalogs/}).


\section*{Acknowledgements}

R.P.S. dedicates this paper to the memory of Prof. Jos\'e Augusto Buarque de Nazareth.  The authors wish to thank workers in the health and services industry who made it possible for this work to be conducted from home during challenging pandemic years. D.M. is supported by ANID BASAL projects ACE210002 and FB210003, and by Fondecyt Project No. 1220724. J.G.F-T gratefully acknowledges the grant support provided by Proyecto Fondecyt Iniciaci\'on No. 11220340, and also from ANID Concurso de Fomento a la Vinculaci\'on Internacional para Instituciones de Investigaci\'on Regionales (Modalidad corta duraci\'on) Proyecto No. FOVI210020, and from the Joint Committee ESO-Government of Chile 2021 (ORP 023/2021), and from Becas Santander Movilidad Internacional Profesores 2022, Banco Santander Chile.
T.C.B. acknowledges partial support from grant PHY 14-30152; Physics Frontier Center/JINA Center for the Evolution of the Elements (JINA-CEE), and from OISE-1927130: The International Research
Network for Nuclear Astrophysics (IReNA), awarded by the US National Science Foundation. 
Funding for the Sloan Digital Sky Survey IV has been provided by
the Alfred P. Sloan Foundation, the U.S. Department of Energy Office
of Science, and the Participating Institutions. SDSS acknowledges
support and resources from the Center for High-Performance Computing
at the University of Utah. The SDSS web site is www.sdss.org. SDSS
is managed by the Astrophysical Research Consortium for the
Participating Institutions of the SDSS Collaboration including the
Brazilian Participation Group, the Carnegie Institution for Science,
Carnegie Mellon University, the chilean Participation Group, the
French Participation Group, Harvard-Smithsonian Center for Astrophysics,
Instituto de Astrof\'{i}sica de Canarias, The Johns Hopkins University,
Kavli Institute for the Physics and Mathematics of the Universe
(IPMU) / University of Tokyo, the Korean Participation Group,
Lawrence Berkeley National Laboratory, Leibniz Institut f\"{u}r Astrophysik
Potsdam (AIP), Max-Planck-Institut f\"{u}r Astronomie (MPIA Heidelberg),
Max-Planck-Institut f\"{u}r Astrophysik (MPA Garching), Max-Planck-Institut
f\"{u}r Extraterrestrische Physik (MPE), National Astronomical Observatories
of china, New Mexico State University, New York University, University
of Notre Dame, Observatório Nacional / MCTI, The Ohio State University,
Pennsylvania State University, Shanghai Astronomical Observatory,
United Kingdom Participation Group, Universidad Nacional Autónoma
de México, University of Arizona, University of Colorado Boulder,
University of Oxford, University of Portsmouth, University of Utah,
University of Virginia, University of Washington, University of
Wisconsin, Vanderbilt University, and Yale University.

This work presents results from the European Space Agency (ESA) space mission Gaia. Gaia data are being processed by the Gaia Data Processing and Analysis Consortium (DPAC). Funding for the DPAC is provided by national institutions, in particular the institutions participating in the Gaia MultiLateral Agreement (MLA). The Gaia mission website is \href{https://www.cosmos.esa.int/gaia}{https://www.cosmos.esa.int/gaia}. The Gaia archive website is \href{https://archives.esac.esa.int/gaia}{https://archives.esac.esa.int/gaia}.

{\it Software:} Astropy \citep{astropy:2013,astropy:2018}, SciPy
\citep{Scipy2020}, NumPy \citep{NumPy}, Matplotlib \citep{Hunter:2007},
Galpy \citep{Galpy2015,Galpy2018}, TOPCAT \citep{Taylor2005}.

{\it Facilities:} Sloan Foundation 2.5m Telescope of Apache Point Observatory (APOGEE-North), Ir\'en\'ee du Pont 2.5m Telescope of Las Campanas Observatory (APOGEE-South), \textit{Gaia} satellite/European Space Agency (\textit{Gaia}).

\section*{Data availability}
All data used in this paper are publicly available at the SDSS-IV DR17 website:
https: /www.sdss.org/dr17/.



\bibliographystyle{mnras}
\bibliography{example} 




\appendix
\label{sec:append}

\section{Comparison with data from the literature}
\label{sec:appendix}

Elemental abundance analysis is a tricky procedure with outputs that depend strongly on a number of factors.  On the empirical side, the results are sensitive to the choice of spectral region as well as the overall quality of the observational data, usually quantified in terms of S/N, resolution, and sampling.  In addition, the adequacy of data reduction methods is critical, with details such as sky subtraction and telluric absorption elimination being particularly relevant in the NIR. 
The outcome is also strongly influenced by the arsenal employed in the analysis, including model atmospheres, line opacities (wavelengths, molecular and atomic excitation/ionization potentials, log~{\it gf}s, damping constants),  spectrum synthesis code, microturbulent velocities, and assumptions such as spherical symmetry versus plane parallel atmospheres, and consideration or not of local thermodynamic equilibrium.  
In modern times, the advent of massive surveys brought to the fore the automation of the core of the spectral analysis, introducing additional uncertainties. It is thus par for the course that the fidelity of any new data set be scrutinised via comparison with numbers generated independently.   

APOGEE data have been regularly contrasted with literature values.  The survey was indeed designed so as to afford such detailed comparisons, which were performed for each successive data release, and published in a number of papers \citep[e.g.,][]{Holtzman2015,Meszaros2015,Holtzman2018,Nataf2019,Jonsson2018,Jonsson2020,Meszaros2020,Meszaros2021}. To our knowledge however, such a detailed examination of DR17 data has not yet been published, particularly within the regime of globular clusters, whose stars inhabit unique loci of chemical composition space.  We briefly examine in this Appendix a few comparisons with data from a large survey and those from other smaller independent studies.

\begin{figure*}	\includegraphics[width=\textwidth]{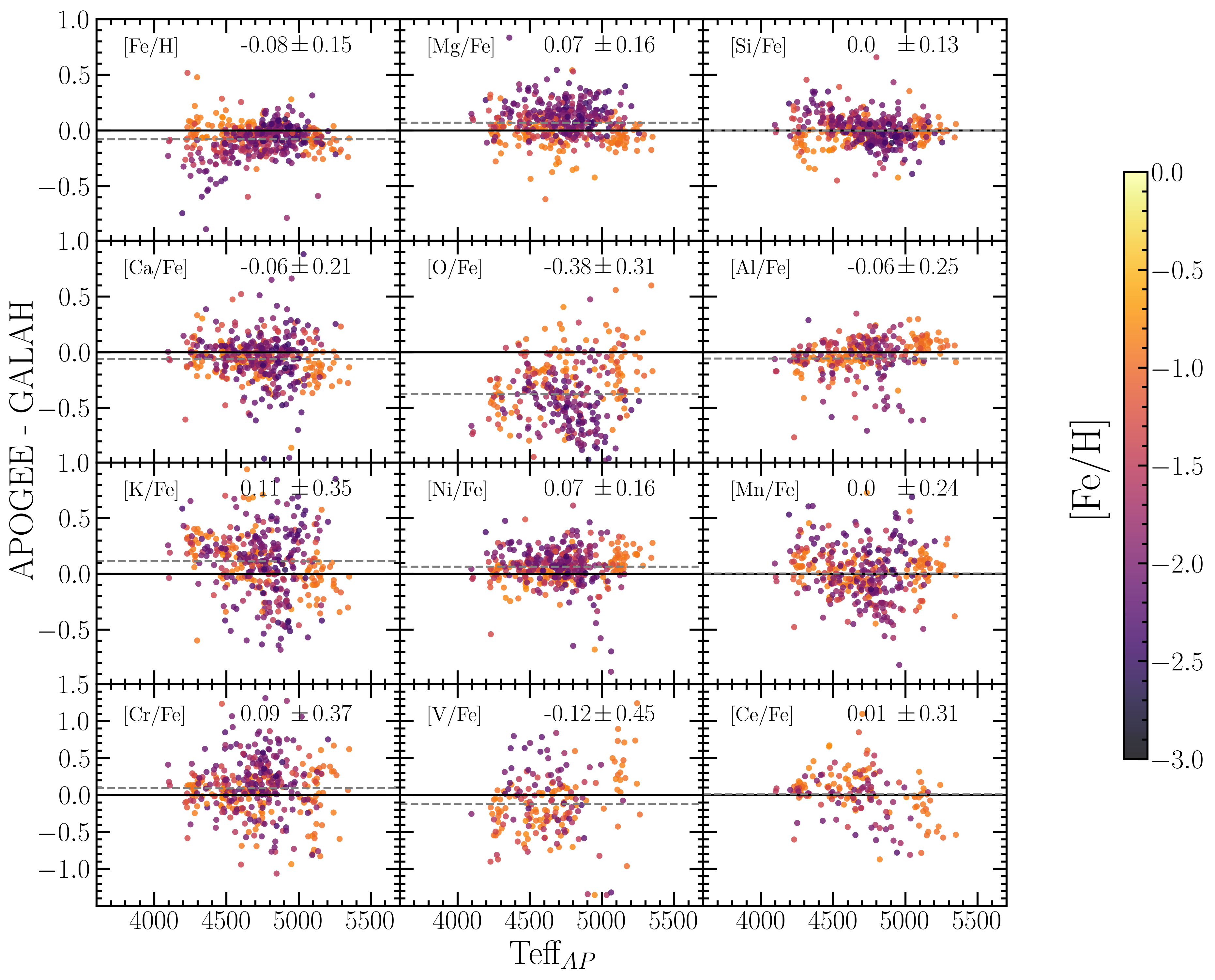}
    \caption{Comparison between elemental abundances for stars in common between  APOGEE and GALAH surveys, colour-coded by [Fe/H]. The solid horizontal lines indicate identical abundances, whereas the dashed gray lines mark the position of the mean residuals.  For all elements, except for oxygen, the mean residuals are much smaller than the r.m.s. of the distribution.}
    \label{fig:apovsgal}
\end{figure*}

We start by comparing our numbers with those generated by the GALAH survey \citep{DeSilva2015,Martell2017}. For that purpose, we matched our sample stars with the GALAH DR3 catalogue \citep{Buder2022}, retaining only the elemental abundances with quality flag=0, which yielded several hundred stars in common for all abundances of interest. 
The comparisons are displayed in Figure~\ref{fig:apovsgal}, and the relevant statistics listed in Table~\ref{tab:residuals}.  Perfect agreement is indicated by the solid black line, whereas the mean difference is marked by the gray dashed line.  Mean residuals and r.m.s. dispersion are indicated on the top right of each panel.  Data points are colour-coded by [Fe/H].
For all abundances the mean residuals are well within the r.m.s., except for the case of oxygen, for which the mean residuals are just above 1~$\sigma$ off.  It is also noteworthy that for some elements, such as O, K, Cr, V, and Ce the dispersion of the abundance ratio residuals is particularly large.  

By looking at the intrinsic dispersion of the abundance ratios in the two data sets, we can pinpoint which of them contributes more importantly to the scatter in the data.  
Columns (3) and (4) of Table~\ref{tab:residuals} display the numbers, obtained by simply calculating the r.m.s. of the abundances from APOGEE and GALAH, using only the stars in common for a fair assessment.  For Mg, Si, O, and Ni the intrinsic scatter in the GALAH data is up to twice larger than APOGEE.  
The opposite is the case for Cr and, to some extent, Ce.  For all the other elements, including those for which a large intrinsic scatter renders the comparison somewhat difficult to interpret (Fe and Al), the two sets have comparable dispersion.  
We conclude that for most elements involved in this comparison, the precision of the APOGEE data is superior to that of GALAH, within this restricted data set.  By the same token, for all elements except oxygen, the zero points of the two abundance systems are indistinguishable from each other.

\begin{table}
\label{tab:residuals}
\centering
\caption{Comparison of abundances from APOGEE and GALAH.  Column information: (1) Iron abundance or abundance ratio; (2) Mean residual and r.m.s. dispersion around the mean; (3) Intrinsic r.m.s. of APOGEE data; (4) Intrinsic r.m.s. of GALAH data; (5) Number of stars in common.}
\begin{tabular}{lrccc} 
\hline
Abundance & Residual & $\sigma_{APO}$ & $\sigma_{GAL}$ & n$_\star$\\
\hline
(1) & (2) & (3) & (4) & (5) \\
\hline
 $\lceil {\rm Fe/H} \rceil$ & $-0.08 \pm 0.15$ & 0.44 & 0.44 & 447  \\
 $\lceil {\rm Mg/Fe} \rceil$ & $0.07 \pm 0.16$ & 0.14 & 0.21 & 423  \\
 $\lceil {\rm Si/Fe} \rceil$ & $0.00 \pm 0.13$ & 0.07 & 0.13 & 441  \\
 $\lceil {\rm Ca/Fe} \rceil$ & $-0.06 \pm 0.21$ & 0.16 & 0.17 & 417  \\
 $\lceil {\rm O/Fe} \rceil$ & $-0.38 \pm 0.31$ & 0.15 & 0.33 & 317  \\
 $\lceil {\rm Al/Fe} \rceil$ & $-0.06 \pm 0.25$ & 0.33 & 0.33 & 294  \\
 $\lceil {\rm K/Fe} \rceil$ & $0.11 \pm 0.35$ & 0.25 & 0.25 & 395  \\
 $\lceil {\rm Ni/Fe} \rceil$ & $0.07 \pm 0.16$ & 0.09 & 0.15 & 398  \\
 $\lceil {\rm Mn/Fe} \rceil$ & $0.00 \pm 0.24$ & 0.20 & 0.18 & 365  \\
 $\lceil {\rm Cr/Fe} \rceil$ & $0.09 \pm 0.37$ & 0.33 & 0.19 & 380  \\
 $\lceil {\rm V/Fe} \rceil$ & $-0.12 \pm 0.45$ & 0.27 & 0.35 & 188  \\
 $\lceil {\rm Ce/Fe} \rceil$ & $0.00 \pm 0.24$ & 0.40 & 0.33 & 165  \\
\hline
\end{tabular}
\end{table}

\begin{figure}	\includegraphics[width=\columnwidth]{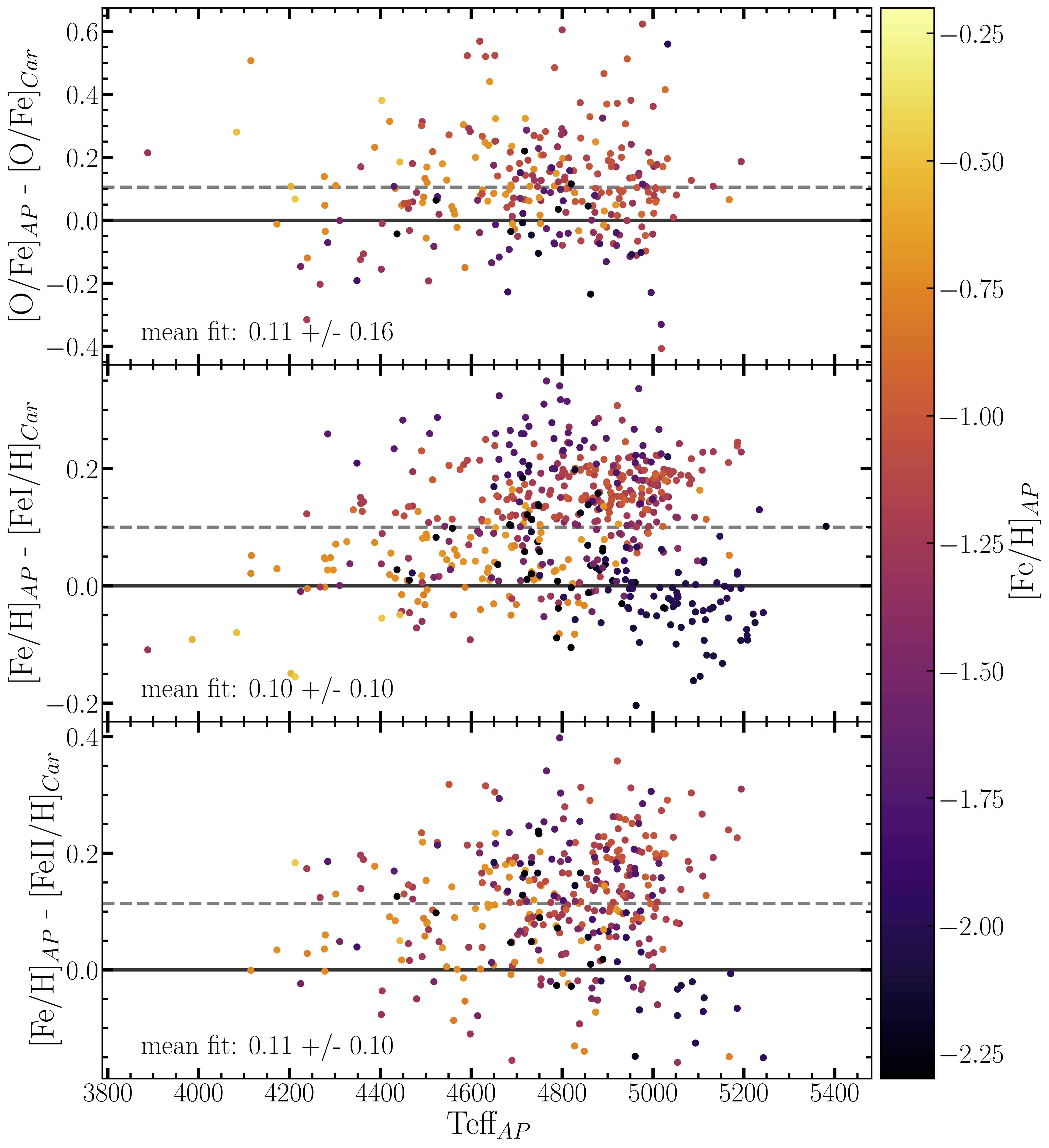}
    \caption{Comparison of APOGEE abundances for O (top panel) and Fe with values obtained by E.~Carretta and collaborators (middle panel for abundances obtained from neutral lines, bottom panel for those based on once-ionised iron).  Good agreements is achieved for [O/Fe], but for [Fe/H] the mean residuals are off at the 1~$\sigma$ level.  There is also evidence for a dependence of residuals on metallicity.}
    \label{fig:apovscar}
\end{figure}

\begin{figure}	\includegraphics[width=\columnwidth]{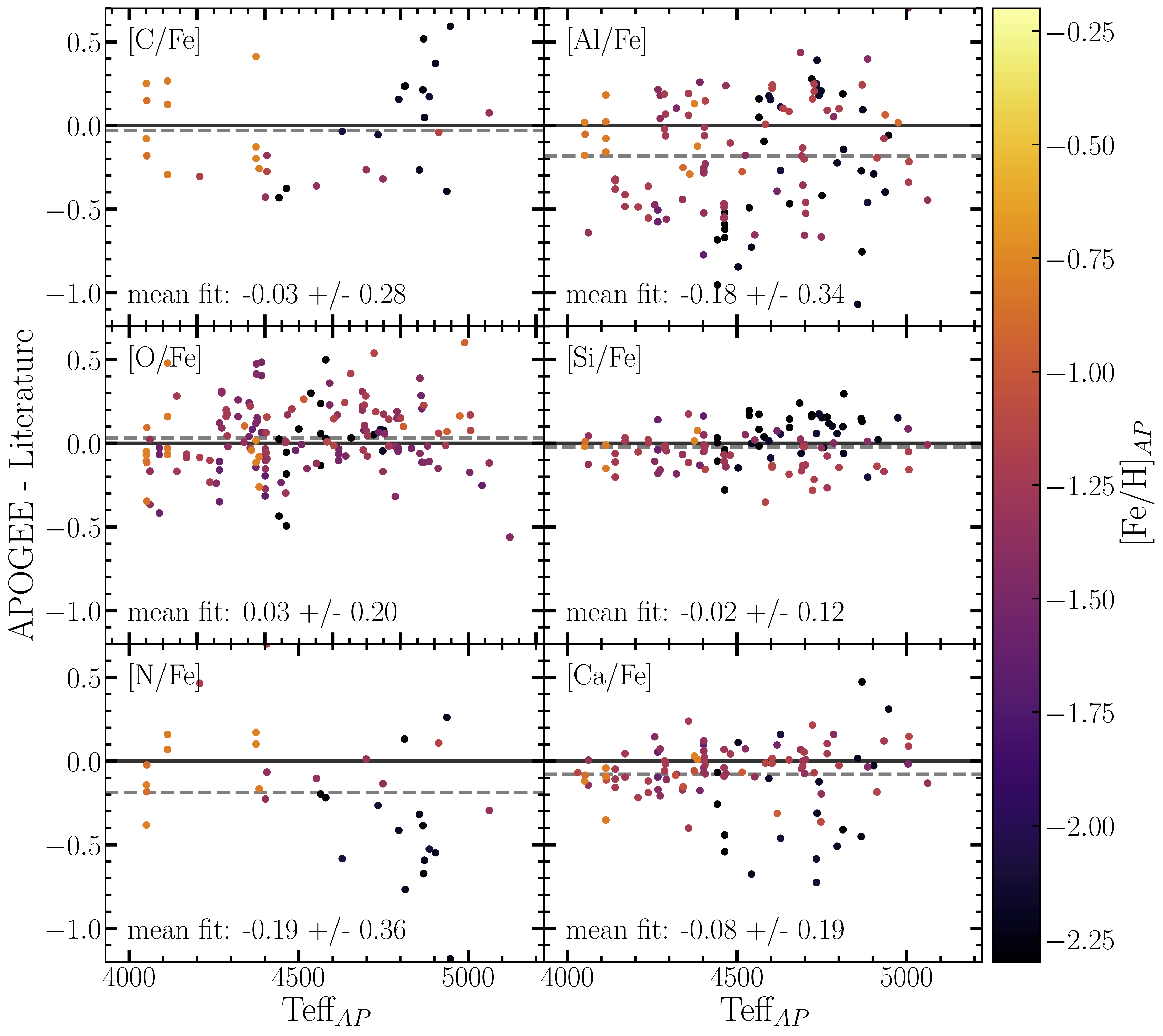}
    \caption{Comparison of APOGEE abundances for other elements with values other sources in the literature.  We find excellent agreement for most elements, and satisfactory agreement for Al and N, with no obvious dependence on metallicity or $T_{\rm eff}$ (except perhaps for N).}
    \label{fig:apovslit}
\end{figure}

To address the matter of data accuracy, we need to resort to comparisons with other literature values based on classical abundance analysis of high-resolution (predominantly optical) spectra.  The best place to start is the extensive data set painstakingly amassed over the years by E.~Carretta and collaborators.  
The compilation presented by \cite{Carretta2010} focuses on Fe, O, and Na, but we limit our discussion to the former two elements.  APOGEE abundances for Na are known to suffer from important shortcomings in the metallicity regime of interest, as they are based on only two lines that are weak in the spectra of metal-poor giants.  Moreover, they are affected by important contamination by airglow emission \citep{Jonsson2020}.  

Stars in common to this value-added catalogue and \cite{Carretta2010} are displayed in abundance  residuals vs. $T_{\rm eff}$ planes in Figure~\ref{fig:apovscar}. The top panel compares APOGEE vs \cite{Carretta2010} values for [O/Fe], whereas comparisons of Fe abundances derived using lines due to neutral and once-ionized iron are displayed in the middle and bottom panels, respectively. Non-negligible differences, at the 1~$\sigma$ level, are seen between the two sets of Fe abundances.  
A mild dependence on metallicity is apparent, with good agreement on the high metallicity end, and a slight deterioration at [Fe/H]$\simless$--1.0. Overall, we conclude that the agreement between the Fe abundances of APOGEE and \cite{Carretta2010} is about satisfactory.
Regarding oxygen abundances,  the mean difference between the two sets is $\sim$0.1~dex, which is well within the r.m.s. scatter, suggesting that the discrepancy between APOGEE and GALAH on the same plane (Figure~\ref{fig:apovsgal}) is due to systematics in the GALAH data.

We wrap up our verification of APOGEE abundances against the literature by extending our scrutiny to additional elements.  Figure~\ref{fig:apovslit} contrast APOGEE abundances with data from a variety of literature sources, originally compiled by \cite{Meszaros2015}\footnote{Sources included are the following: \cite{Briley1997}, \cite{Carretta2009}, \cite{Cavallo2000},
\cite{Cohen2005}, \cite{Ivans2001}, \cite{Johnson2005,Johnson2012}, \cite{Koch2010}, \cite{Kraft1992,Kraft2003},  \cite{Lai2011},
\cite{Lee2004}, \cite{Melendez2009}, \cite{Minniti1996},
\cite{OConnell2011}, \cite{Otsuki2006}, \cite{Ramirez2002,Ramirez2003},
 \cite{Roederer2011}, \cite{Shetrone1996},
\cite{Smith2007}, \cite{Sneden1991,Sneden1992,Sneden1997,Sneden2000,Sneden2004}, \cite{Sobeck2011}, \cite{Yong2006a,Yong2006b},
and \cite{Yong2008}}.  
Abundances shown are those for C, N, O, Ca, Si, and Al.  For all elements, the mean residuals are consistent with with the identity line well within the r.m.s. dispersion.  For N and Al the scatter is larger, and in the case of the former there seems to be a dependence on metallicity.  
This is perhaps not surprising, because APOGEE N abundances rely on CN lines, which become vanishingly weak in giant stars with [Fe/H]$\simless$--2.0.  Outside that regime, agreement between APOGEE and the literature sources is actually very good.

In conclusion, comparison between APOGEE chemical compositions and data from GALAH and a compilation of classical abundance analyses from the literature indicates that APOGEE chemistry for Galactic globular cluster members is characterised by excellent precision and very good accuracy.



\bsp	
\label{lastpage}
\end{document}